\documentclass{aa}
\usepackage{natbib}
\usepackage{graphicx}
\usepackage{epsfig}
\usepackage{amsmath}
\usepackage{txfonts}
\usepackage{hyperref}
\usepackage{url}
\usepackage{xcolor}
\hypersetup{
colorlinks,
linkcolor={red!80!black},
citecolor={blue!80!black},
urlcolor={blue!80!black}
}

\def\caiidoub{[Ca\two]\,$\lambda\lambda$\,$7291,\,7323$}

\def\cm3{cm$^{-3}$}
\def\kms{km~s$^{-1}$}

\def\lsun{L$_{\odot}$}
\def\rsun{R$_{\odot}$}

\def\msun{M$_{\odot}$}

\def\two{\ts {\,\sc ii}}

\def\beq{\begin{equation}}
\def\eeq{\end{equation}}

\def\lesssim{\mathrel{\hbox{\rlap{\hbox{\lower4pt\hbox{$\sim$}}}\hbox{$<$}}}}
\def\gtrsim{\mathrel{\hbox{\rlap{\hbox{\lower4pt\hbox{$\sim$}}}\hbox{$>$}}}}

\def\isoni{$^{56}{\rm Ni}$}

\def\two{{\,\sc ii}}

\def\ip{\rho}

\newcommand{\code}[1]{\texttt{#1}}

\def\voned{{\code{V1D}}}

\def\mesa{{\code{MESA}}}
\def\cmfgen{{\code{CMFGEN}}}

\def\longpol{{\code{LONG\_POL}}}

\newcommand{\iso}[2]{\ensuremath{^{#1}\rm{#2}}}

\def\pasp{PASP}

\def\apj{ApJ}
\def\apjs{ApJS}
\def\apjl{ApJL}
\def\aap{A\&A}
\def\araa{ARA\&A}

\def\mnras{MNRAS}
\def\nat{Nature}

\def\nifs{\iso{56}Ni}
\def\cofs{\iso{56}Co}

\begin{document}

   \title{The evolution of continuum polarization in Type II supernovae as a diagnostic of ejecta morphology}
   \titlerunning{Continuum polarization of Type II SNe}

\author{
   Luc Dessart\inst{\ref{inst1}}
  \and
   D. John Hillier\inst{\ref{inst2}}
  \and
   Douglas C. Leonard\inst{\ref{inst3}}
}

\institute{
Institut d'Astrophysique de Paris, CNRS-Sorbonne Universit\'e,  98 bis boulevard Arago, F-75014 Paris, France.\label{inst1}
  \and
    Department of Physics and Astronomy \& Pittsburgh Particle Physics, Astrophysics, and Cosmology Center (PITT PACC),  University of Pittsburgh, 3941 O'Hara Street, Pittsburgh, PA 15260, USA.\label{inst2}
  \and
    Department of Astronomy, San Diego State University, San Diego, CA 92182-1221, USA.\label{inst3}
}

   \date{Received; accepted}
  \abstract{Linear polarization of the optical continuum of type II supernovae (SNe), together with its temporal evolution, is a promising source of information on the large-scale geometry of their ejecta. To help tap this information we have undertaken 2D polarized radiative transfer calculations to map out the possible landscape of type II SN continuum polarization ($P_{\rm cont}$) from 20 to 300\,d after explosion. Our simulations are based on crafted 2D, axisymmetric ejecta constructed from 1D nonlocal thermodynamic equilibrium time-dependent radiative transfer calculations for a red-supergiant star explosion. Following the approach used in our previous work on SN\,2012aw, we consider a variety of bipolar explosions in which spherical symmetry is broken by the presence, within $\sim$\,30$^\circ$ of the poles, of material with a higher kinetic energy (up to a factor of two) and higher \nifs\ abundance (up to a factor of about five, with allowance for \nifs\ at high velocity). Our set of eight 2D ejecta configurations produces considerable diversity in $P_{\rm cont}$ ($\lambda \sim 7000$\,\AA), although its maximum of 1--4\,\% occurs systematically around the transition to the nebular phase. Before and after that transition, $P_{\rm cont}$ may be null, constant, rising, or decreasing, which results from the complex geometry of the depth-dependent density and ionization as well as  from optical depth effects. Our modest angle-dependent explosion energy can yield $P_{\rm cont}$ of 0.5--1\,\% at early times. Residual optical-depth effects can yield an angle-dependent SN brightness and constant polarization at nebular times. Observed values of $P_{\rm cont}$ tend to be lower than obtained here, suggesting more complicated geometries with competing large-scale structures causing polarization cancellation. Extreme asymmetries seem to be excluded.
}

\keywords{
  radiative transfer --
  polarization --
  supernovae: general
}
   \maketitle

\section{Introduction}
\label{sect_intro}

A fundamental property of the neutrino-driven explosion mechanism of massive stars is its inherently multidimensional character (see, for example, \citealt{bollig_ccsn_3d_21}; \citealt{mezzacappa_ccsn_20}; \citealt{oconnor_couch_3d_18}; \citealt{vartanyan_ccsn_22}). Although such asymmetry may leave an imprint on the SN light curve and spectra, its most unambiguous signature in non-spatially resolved ejecta is a residual, nonzero linear polarization  \citep{shapiro_sutherland_82,wang_wheeler_rev_08}.

Polarization studies of SNe started in earnest with the Type II-peculiar SN\,1987A, with multi-band polarimetry (e.g., \citealt{mendez_87A_pol_88}) as well as spectropolarimetry (e.g., \citealt{cropper_specpol_87a_88}). \citet{hoeflich_87A_91} and \citet{jeffery_87_pol_91} modeled the intrinsic linear polarization of SN\,1987A and proposed that it originates from electron scattering in an asymmetric ejecta -- the polarization is understood as arising from an aspherical scattering photosphere. \citet{chugai_pol_87a_92} proposed that an asymmetric distribution of \nifs\ could also be the source of the SN\,1987A polarization. Because of the scarcity of Type II-peculiar SNe like 1987A amongst non-interacting Type II SNe, recent spectropolarimetric observations have instead primarily been gathered for Type II-Plateau (II-P) SNe. The first complete coverage of the spectropolarimetric evolution of a Type II-P SN was for 2004dj \citep{leonard_04dj_06}, which revealed a low intrinsic continuum polarization through the plateau phase, which then spiked at the onset of the nebular phase, and followed a $1/t^2$ drop subsequently. While this naively appeared to imply a greater asymmetry of the inner ejecta relative to the outer ejecta, \citet{DH11_pol} found that the polarization peak is in part driven by the transition from a multiple- to a single-scattering regime, hence a radiative transfer effect -- this was later confirmed with the simulations of \citet{dessart_12aw_21} for SN\,2012aw with more consistent models covering the full evolution of a 2D ejecta from the photospheric phase to the nebular phase. This behavior is however not generic since SNe II-P exhibit a variety of behavior (e.g., \citealt{chornock_pol_10,leonard_12aw_12}), with small polarization at all times (e.g., SN\,1999em; \citealt{leonard_99em_specpol_01}), a polarization peak before the end of the plateau phase (e.g., SN\,2013ej \citealt{leonard_iauga_15}; \citealt{mauerhan_13ej_17}; \citealt{nagao_13ej_21}; see also SN\,2017gmr; \citealt{nagao_17gmr_pol_19}), or a near constant polarization at nebular times (e.g., SNe 2008bk or 2013ej; \citealt{leonard_08bk_12,leonard_iauga_15}).

The notion that the inner ejecta of Type II SNe is asymmetric is indirectly supported by the detection of nonzero polarization in stripped envelope SNe -- objects in which there is no massive H-rich envelope to inhibit the expansion of the metal-rich asymmetric core. Although the absence of hydrogen reduces the importance of electron scattering in favor of line opacity and therefore complicates the interpretation of the polarization signatures, intrinsic linear polarization has been routinely detected in Type IIb SNe (e.g., SN\,1993J; \citealt{trammell_93J_93}; SN\,2008ax; \citealt{chornock_etal_11}), Type Ib SNe (e.g., SN\,2008D; \citealt{maund_08D_09}), Type Ic SNe (e.g., SN\,2007gr; \citealt{tanaka_07gr_specpol_08}), and even broad-lined Type Ic SNe (e.g., SN\,2002ap;  \citealt{leonard_02ap_pol_02}).

Such polarization measures suggest an asymmetry of the explosion causing an asymmetric density distribution \citep{hoeflich_87A_91,jeffery_87_pol_91,DH11_pol} or an asymmetric distribution of the \nifs\ \citep{chugai_04dj_06,dessart_pol_blob_21,leonard_13ej_21}. Because the \nifs\ mass scales with the explosion energy,\footnote{The positive correlation between \nifs\ mass and explosion energy is inferred from observations (see, e.g., \citealt{pejcha_prieto_sn2p_15}) and is theoretically understood as arising from the dependence between explosion energy, post-shock temperature, and explosive nucleosynthesis. Roughly speaking, shocked material with a temperature in excess of $5 \times 10^9$\,K burns to iron-group elements. If its electron fraction is close to or equal to 0.5, the main isotope will be \nifs\ (see, for example, Section\,{\sc viii} of  \citet{whw02}.} the ejecta asymmetry is probably caused by a combination of both effects. \citet{DH11_pol} emphasized that an asymmetric density distribution does not just impact the distribution of the free electrons that scatter the radiation but also the distribution of the escaping flux as observed on the plane of the sky. Since radiation tends to escape from regions of lower optical depth, an oblate distribution of free electrons may induce a prolate distribution of the escaping flux and produce a complicated polarization signature. A globally consistent model of the polarization is therefore necessary, especially at times when the ejecta are optically thick. Hence, in order to improve the physical realism of the initial conditions for these spectropolarimetric modeling studies, \citet{dessart_12aw_21} set up a 2D axially-symmetric ejecta using physically-consistent 1D nonlocal thermodynamic equilibrium (non-LTE) radiative transfer calculations computed with \cmfgen\ \citep{HD12} and started from radiation-hydrodynamics models of the explosion. Post-processing with the 2D polarized radiative transfer code \longpol\ \citep{hillier_94,hillier_96,DH11_pol} shows that an asymmetric explosion of a standard Type II-P SN may explain the photometric, spectroscopic, and multiwavelength polarization evolution of a SN like 2012aw \citep{dessart_12aw_21}.

In this study, we consider a broad variety of 2D axisymmetric but aspherical ejecta and explore the evolution in the continuum polarization from early times in the photospheric phase until  300\,d in the nebular phase. We consider variations in kinetic energy and \nifs\ mass with angle, including various combinations thereof, which therefore imply radial variations in both density and \nifs\ mass fraction with angle. In particular, we aim to address the origin of large polarization at early epochs, the range of polarization peaks attained, the behavior at late times, and importantly, understand what drives this diversity. In the next section, we present our numerical setup, both for the 1D \cmfgen\ simulations and the 2D \longpol\ simulations. The grid of models is composed of ejecta with a range of kinetic energies and \nifs\ mass, invoking the presence of  \nifs\ at low or high velocity. In Section~\ref{sect_pcont_ref}, we describe in detail the results for a representative model. In Section~\ref{sect_pcont_all}, we describe the results for the full set of 2D models, addressing in turn the physical conditions that produce the different evolution of the continuum polarization. We present our conclusions in Section~\ref{sect_conc}.\footnote{An observational study with a similar focus but different methodology was published by \citet{nagao_pcont_23} as we completed the writing of this manuscript. The bulk of the work and analysis presented here was done two years ago.}

\section{Numerical setup}
\label{sect_setup}

  The calculations presented in this work are performed in two steps. We first generate 1D explosion models for which we compute the evolution until 300\,d with \cmfgen\ (Section~\ref{sect_1d}). The second step is to build 2D axisymmetric ejecta by combining different pairs of 1D ejecta. For each of these pairs, we compute the 2D polarized radiative transfer with \longpol\ (Section~\ref{sect_2d}) and extract the continuum polarization at multiple epochs. Since a separate project is to analyze our VLT-FORS spectropolarimetric data for SN\,2008bk (see preliminary results in \citealt{leonard_08bk_12,leonard_iauga_15}), we chose for this work a 12\,\msun\ progenitor producing a low energy explosion \citep{lisakov_08bk_17}. This way, the present set of simulations may also be used for the analysis of the SN\,2008bk spectropolarimetry.

\subsection{Spherically-symmetric calculations with \cmfgen}
\label{sect_1d}

 All simulations in this work are based on a nonrotating 12\,\msun\ star initially and evolved at solar metallicity\footnote{So far, spectropolarimetric observations have typically been obtained for nearby events, which for the vast majority arise from solar-metallicity environments, so this choice is sensible.  In practice, metal-line blanketing reduces the albedo and has an adverse effect on polarization. Variations in metallicity will modulate this. However, when considering spectral regions that are relatively line free at most photospheric epochs (around 7000\,\AA), metallicity variations should have little impact. Once at the nebular phase, the continuum is weak and lines dominate, many arising from metals produced during the explosion (hence, unaffected by the original metallicity of the progenitor on the zero-age main sequence).} until core collapse (i.e., when the maximum Fe-core infall velocity exceeds 1000\,\kms) with the code \mesa\ version 10108 \citep{mesa1,mesa2,mesa3,mesa4}. The default parameters are used apart from a reduced wind mass loss rate relative to the Dutch recipe (we adopt a scaling of 0.6). As in our previous works (see, e.g.,  \citealt{d13_sn2p}), we  increase the mixing length parameter to three to produce a relatively compact red-supergiant star at collapse. At core collapse, the model has a luminosity of 56700\,\lsun, a surface radius of 485\,\rsun, an effective temperature of 4040\,K, a total mass of 10.34\,\msun, an H-rich envelope mass of a 7.03\,\msun, a He core mass of 3.31\,\msun, and an Fe core mass of 1.5\,\msun\ (as defined by the innermost envelope location where the electron fraction drops below 0.499).

This progenitor model is exploded with \voned\ \citep{livne_93,dlw10a,dlw10b} by depositing energy for 500\,ms in the innermost 0.05\,\msun\ above an adopted mass cut at a Lagrangian mass of 1.58\,\msun, which corresponds to the location where the entropy rises to 4\,$k_{\rm B}$\,baryon$^{-1}$. The deposited energy equals the binding energy of the overlying envelope (which is $-1.7 \times 10^{50}$\,erg) plus an extra energy of 2 or $4 \times 10^{50}$\,erg (model series named e1 and e2 ). Some \nifs\ is produced during the explosion but for our controlled experiment, it was more practical to reset this \nifs\ mass. Once the explosive nucleosynthesis is over (at about 1$-$2\,s after the explosive trigger), we scale the \nifs\ mass fraction profile to match a specific  \nifs\ mass. This isotope is subdominant (on the order of a few 0.01\,\msun\ in a $\sim$\,10\,\msun\ ejecta) so this has little impact on other species. Once the \nifs\ mass fraction is adjusted, we scale all other mass fractions for a normalization to unity (i.e., we enforce at each depth $\sum_i X_i = 1$). Our 1D explosion models were thus reset to have a \nifs\ mass of 0.009 or 0.05\,\msun\ (suffix ni1 and ni2). Furthermore, when resetting the \nifs\ mass of some models (suffix b1 and b2), we also added an outer \nifs-rich shell of 0.02 or 0.05\,\msun\ (the shell profile is taken to be a gaussian  with a center at 8\,\msun\ and a characteristic width of 1\,\msun). In models with suffix b1 or b2 but characterized by different explosion energies, this outer \nifs-rich shell is located at different velocities. The full set of 1D models that we produced are e1ni1, e1ni2, e1ni1b1, e1ni1b2, e2ni1, e2ni2, e2ni1b1, e2ni1b2. A summary of the properties of these 1D models is presented in Table~\ref{tab_set}.

Adjusting the \nifs\ abundance profile is a common approach in 1D simulations. While artificial, the approach is suitable for testing the influence of  the \isoni\  abundance profile on observations, and has been justified in Section~2 of \citet{dessart_pol_blob_21}. We have used this approach in both \citet{dessart_pol_blob_21,dessart_12aw_21} and \citet{leonard_13ej_21}. The choice of a large \nifs\ excess is made to yield unambiguous signatures that can be clearly identified, while the choice of a Lagrangian mass of 8\,\msun\ was made to place the excess \nifs\ at a few 1000\,\kms\ since such velocities are both predicted in 3D neutrino-driven explosions (e.g., \citealt{gabler_3dsn_21}), as well as inferred from observations (e.g., SN\,2012aw, \citealt{dessart_pol_blob_21,dessart_12aw_21}; and SN\,2013ej, \citealt{leonard_13ej_21}). 

\begin{figure}[t]
\epsfig{file=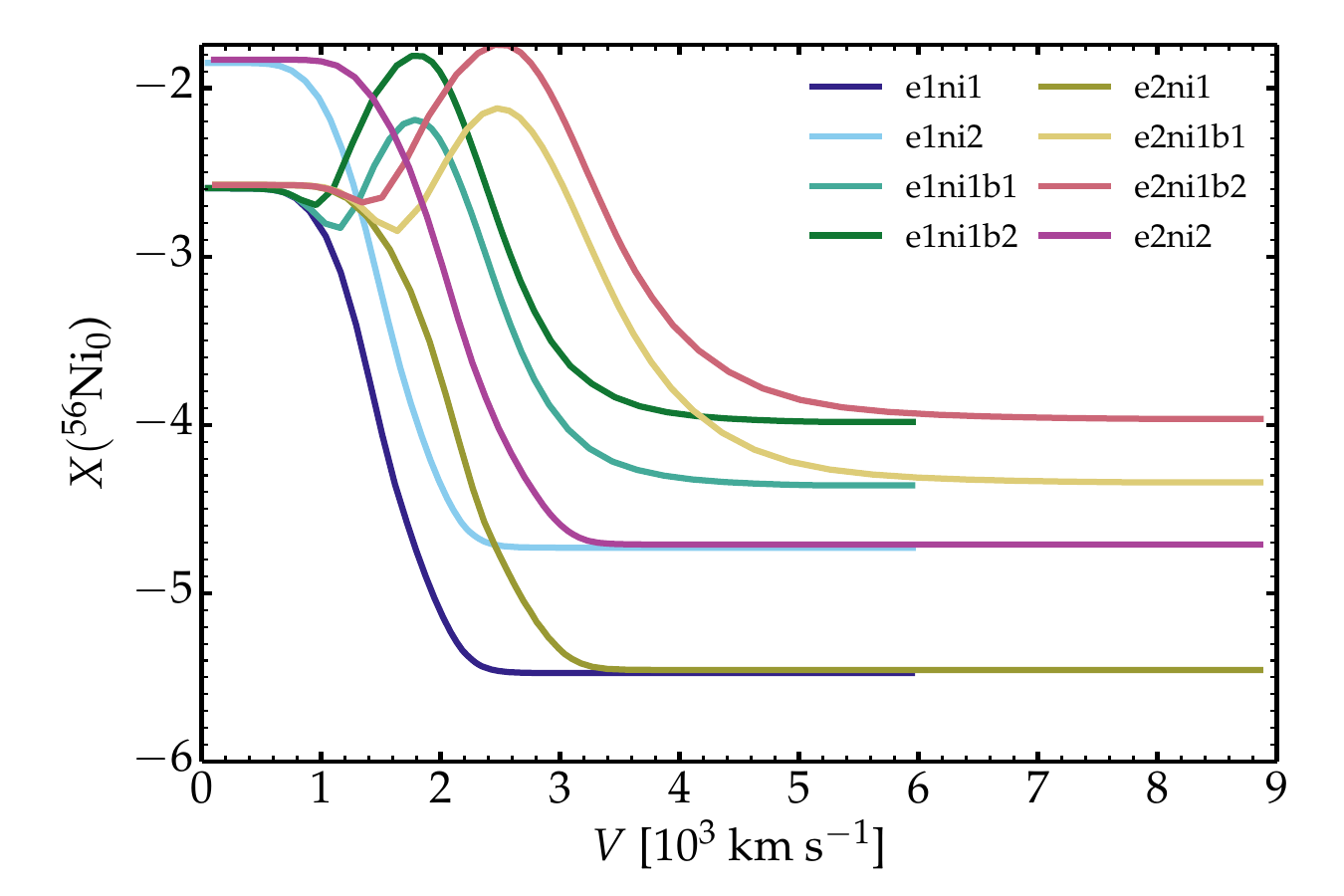, width=9.2cm}
\epsfig{file=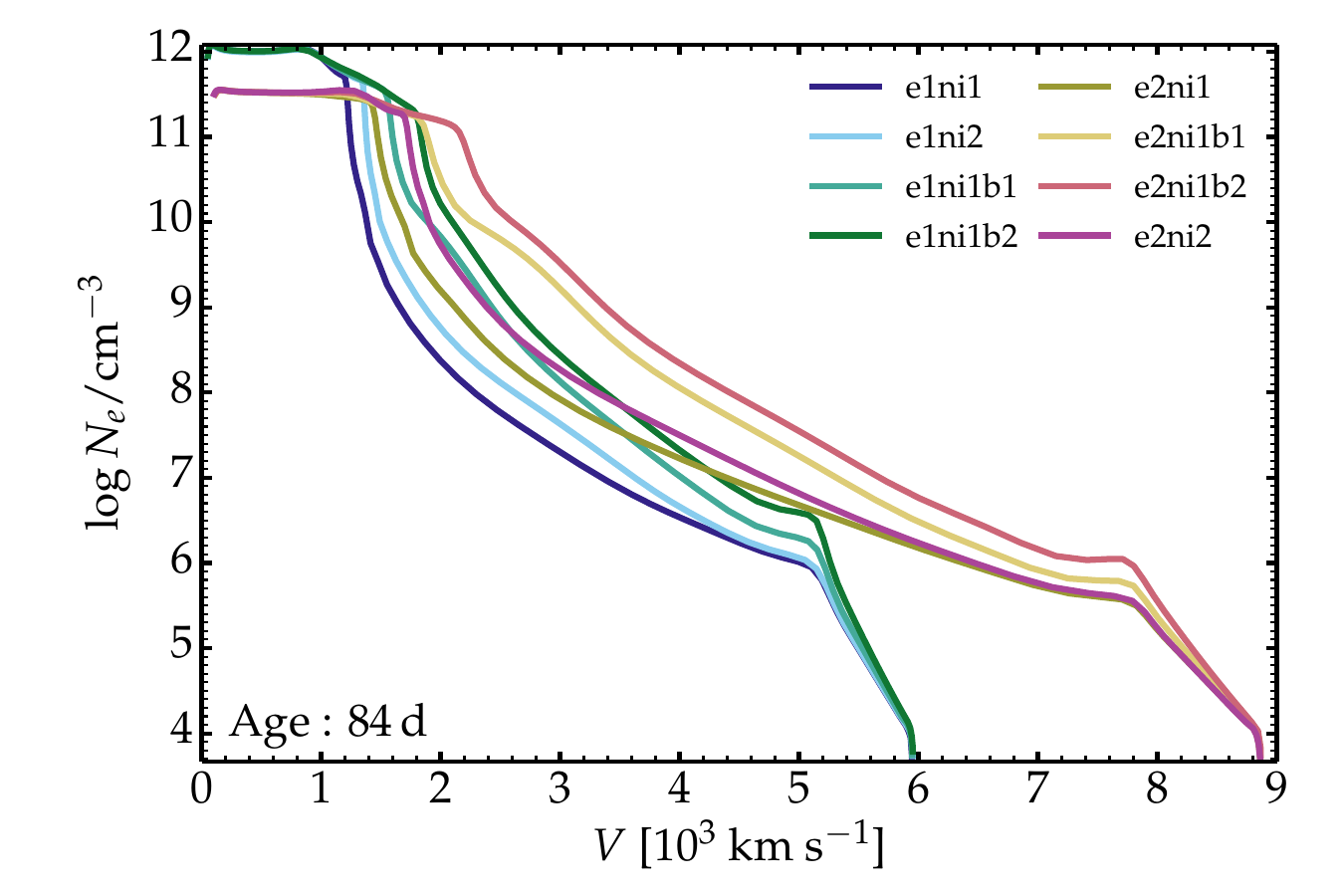, width=9.2cm}
\caption{Ejecta properties for our 1D \cmfgen\ model set. Top: profile of the undecayed \nifs\ versus velocity. The ``bump'' in \nifs\ abundance at higher velocities for  models labelled by a ``b1/2'' suffix  does not appear in the other models in which there is only a central concentration of \nifs. Bottom: profile of the electron density  versus velocity at 84\,d after explosion.
  \label{fig_1d_ejecta}
}
\end{figure}

\begin{figure}
\epsfig{file=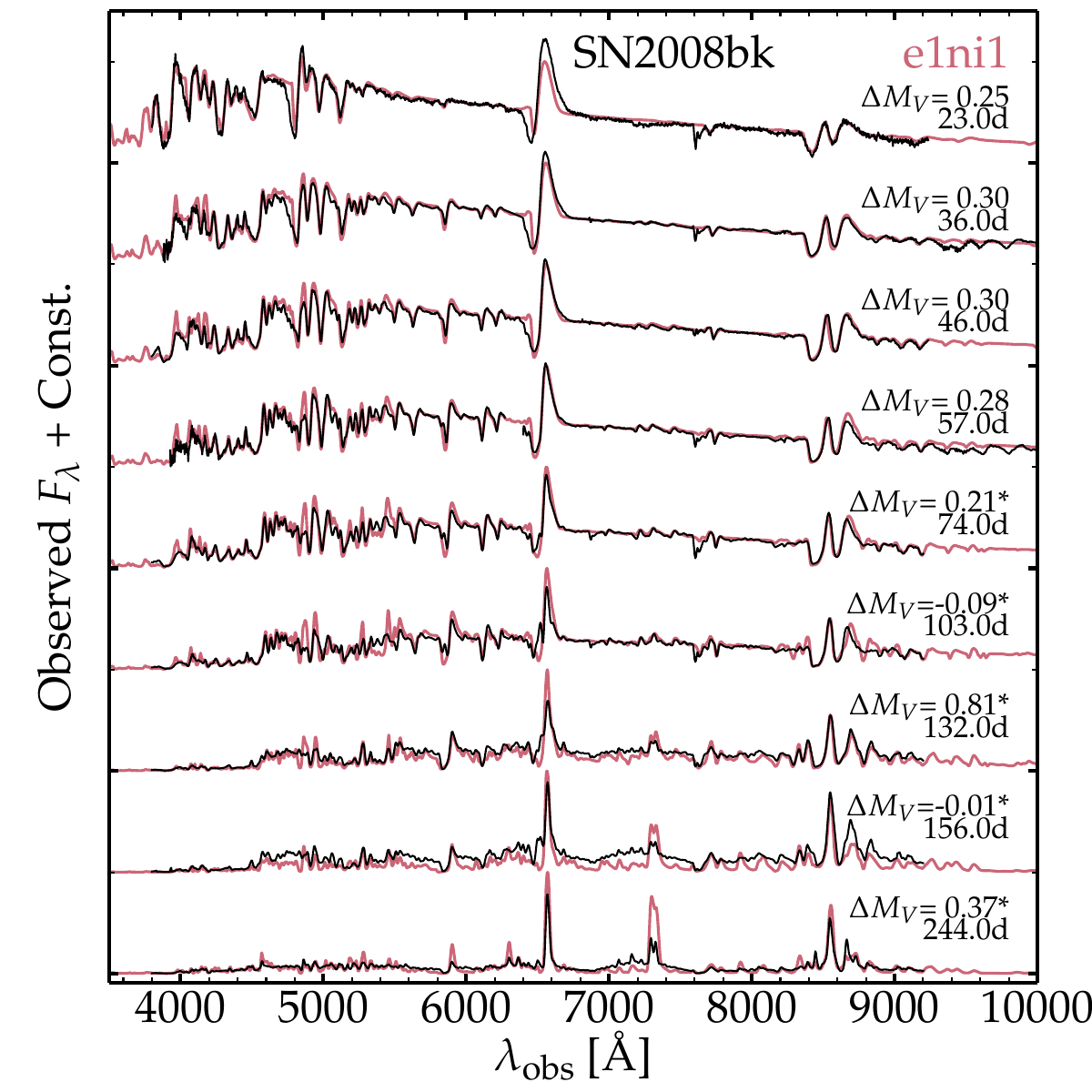, width=9.2cm}
\caption{Spectral comparison between model e1ni1 and the observations of SN\,2008bk after correction for redshift and reddening \citep{lisakov_08bk_17}. Epochs with a starred label correspond to data from \citet{leonard_08bk_12} and the others are from \citet{pignata_08bk_13}.
  \label{fig_spec_1d_08bk}
}
\end{figure}

\begin{figure}
\epsfig{file=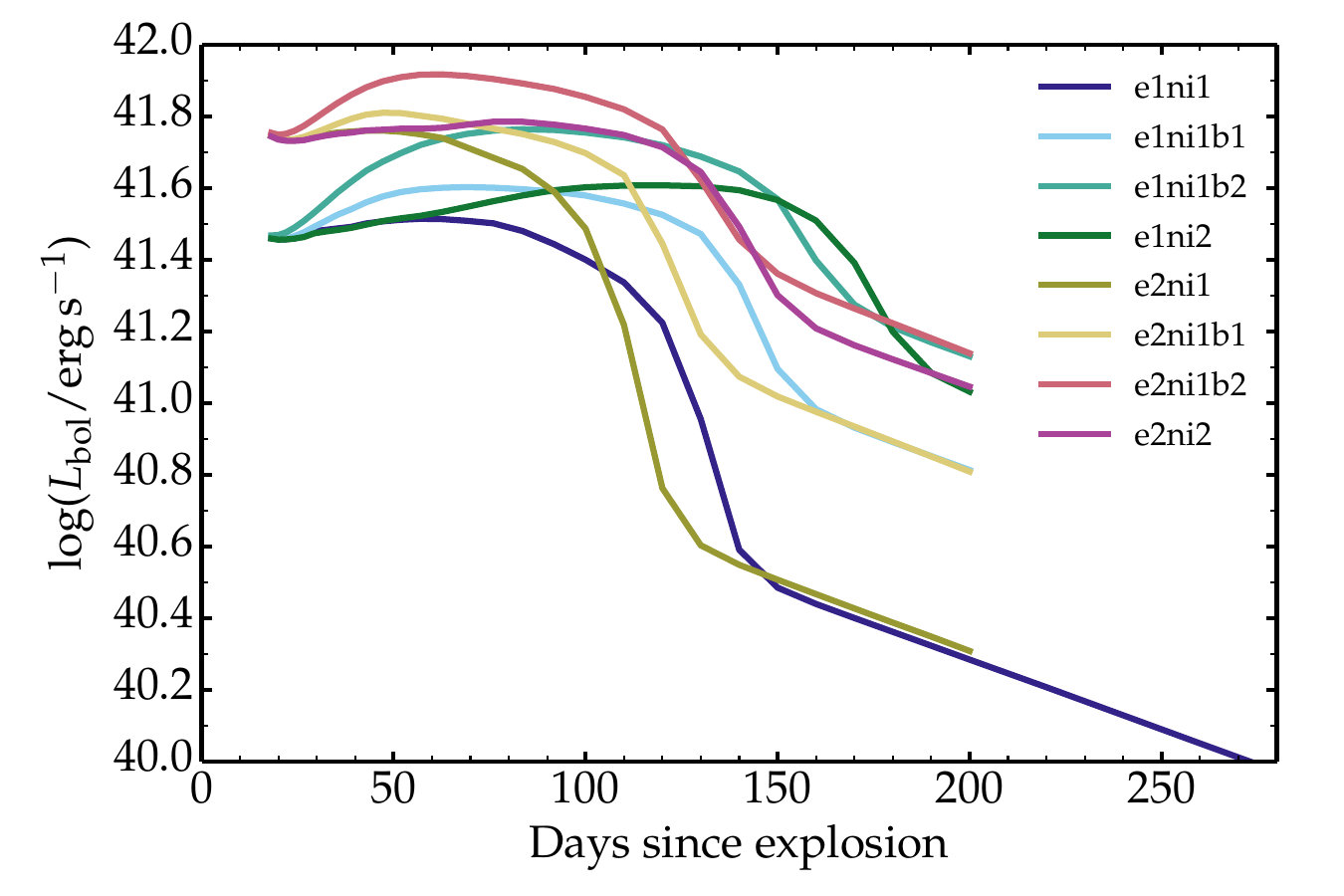, width=9.2cm}
\caption{Model light curves for our set of 1D \cmfgen\ simulations.
  \label{fig_1d_lc}
}
\end{figure}

A fundamental ingredient leading to linear polarization of Type II-P SN radiation is the distribution of free electrons, which is controlled by numerous non-LTE processes as well as by time-dependent effects \citep{UC05,D08_time}, non-thermal processes \citep{lucy_91,swartz_ib_91,li_etal_12_nonte,d12_snibc}, or the composition (H-rich versus H-poor). Starting from these explosions produced with \voned, we compute 1D \cmfgen\ time sequences  from 15\,d\footnote{\cmfgen\ solves for the radiative transfer but ignores the hydrodynamics. So, we must start the \cmfgen\ sequences when the dynamical phase is over, which takes 10$-$15\,d in RSG star explosions because of the slow reverse shock progressing inward into the metal rich ejecta (see discussion in \citealt{DH11_2p}).} until 300\,d. These calculations are done using the standard technique (see, for example, \citealt{HD19} for details) and allow us to compute the evolution of the ejecta properties (including the electron density versus velocity) as well as  the radiation properties. Figure~\ref{fig_1d_ejecta} shows the initial \nifs\ composition profile for the 1D model set as well as the corresponding free electron density at 84\,d after explosion. Evidently, the different explosion energies and \nifs\ yields have a clear impact. A larger explosion energy has more mass at large velocity, which increases the density of free electrons as long as the ionization remains comparable. A greater \nifs\ abundance also leads to enhanced heating and non-thermal ionization, which boost the free-electron density. The effect of the outer, ``shell'' region of enhanced \nifs\ is to increase the free-electron density throughout the outer ejecta, not just in a confined region surrounding the shell itself.

 We show in Fig.~\ref{fig_spec_1d_08bk} the spectral evolution for model e1ni1 together with the contemporaneous spectra of SN\,2008bk. As expected, the agreement is good and similar to that obtained by \citet{lisakov_08bk_17} since  progenitors and explosion characteristics are similar in both works. In the appendix, Fig.~\ref{fig_spec_1d_comp} illustrates the spectral differences at 20 and 84\,d after explosion between these 1D models computed with \cmfgen. Models with twice greater ejecta kinetic energies show broader lines early on but the trend may break down at later times in the photospheric phase since its duration varies between models (for example, all else being the same, a greater ejecta kinetic energy shortens the plateau phase), as evident from the bolometric light curves (Fig.~\ref{fig_1d_lc}).  Models with more \nifs\ show a rising light curve in the second half of the photospheric phase, transition to the nebular phase later, and because of full $\gamma$-ray  trapping stay brighter in the nebular phase.

\begin{figure}
\epsfig{file=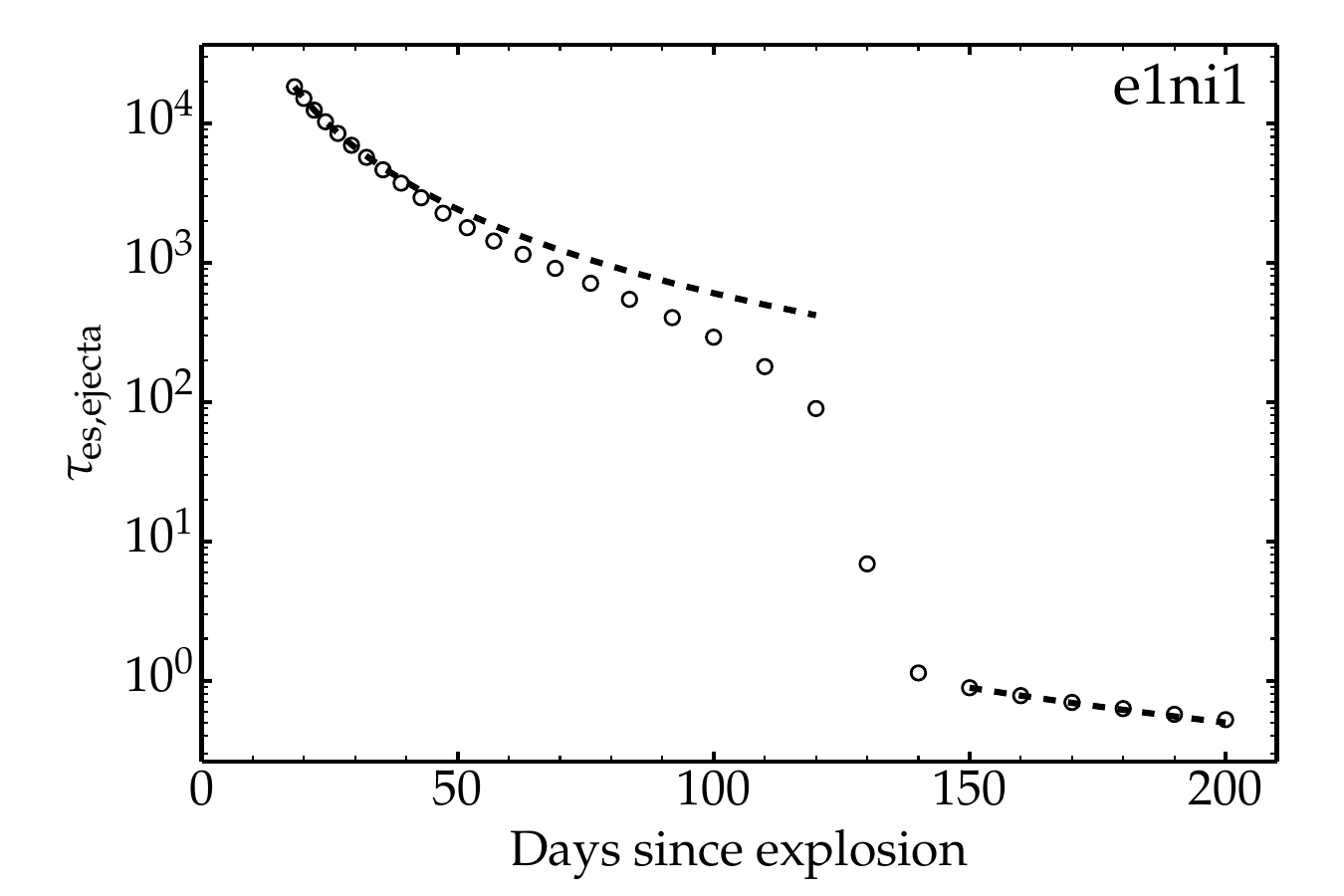, width=9.2cm}
\caption{Evolution of the total ejecta electron-scattering optical depth for model e1ni1. The two dashed curves give the expected $1/t^2$ evolution for an ejecta with a constant ionization.\label{fig_tau_es_e1ni1}
}
\end{figure}

The difference in electron density profiles between models translates into a different evolution of the ejecta electron-scattering optical depth $\tau_{\rm es}$. The time when $\tau_{\rm es}$ drops to one can be extracted from the ejecta properties directly, but it can also be inferred from the light curve (Fig.~\ref{fig_1d_lc}) since it corresponds to the time when the bolometric luminosity falls onto the nebular tail. This time varies between 130\,d (model e2ni1; relatively low \nifs\ mass but higher kinetic energy) and 190\,d (e1ni2; relatively high \nifs\ mass but lower kinetic energy). As an example, Fig.~\ref{fig_tau_es_e1ni1} shows the evolution of the total ejecta electron-scattering optical depth for model e1ni1. The jump at about 130\,d corresponds to the end of the plateau phase when the photosphere rapidly recedes through the metal-rich core. Before and after, the curve follows closely the $1/t^2$ evolution expected for constant ejecta ionization due to geometrical dilution (for discussion, see \citealt{DH11_2p}).

\begin{figure}
\epsfig{file=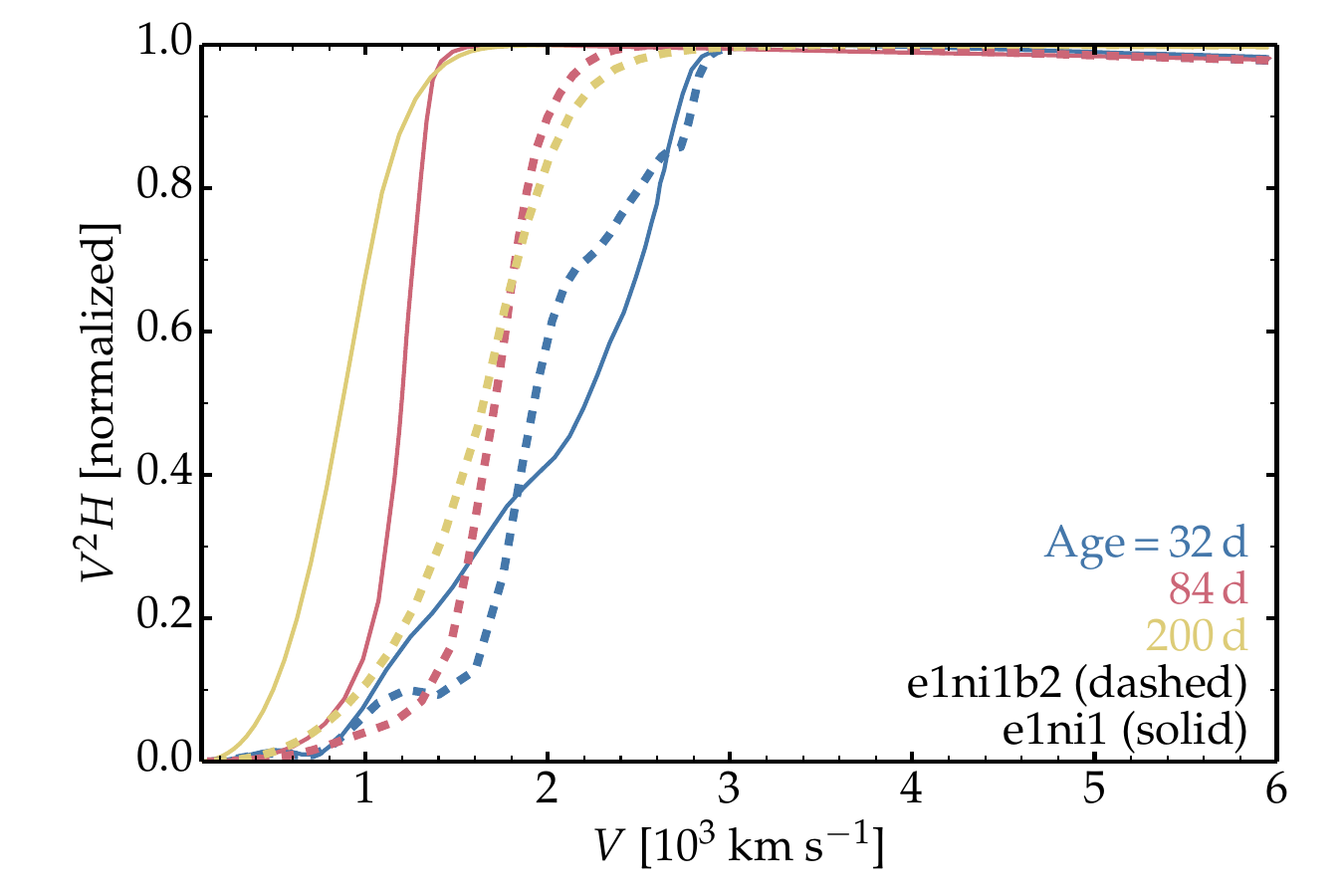, width=9.2cm}
\caption{Variation of the comoving-frame bolometric flux $H$ (scaled by $V^2$ and normalized) versus velocity for models e1ni1 (solid) and e1ni1b2 (dashed) at three epochs covering the early photospheric phase (SN age of 32\,d; blue), the recombination epoch (84\,d; red), and the nebular phase (200\,d; yellow). The radiation from the SN ejecta does not arise from a localized central point source, nor from a well-defined narrow layer, such as the photosphere. Instead the SN flux forms, at all times, over a range of ejecta depths.
\label{fig_flux_vs_vel}
}
\end{figure}

Finally, polarization studies often idealize the emitting source of radiation as a point source, as in the case of a star illuminating an optically thin nebula \citep{Brown_McLean_77}. As discussed in \citet{DH11_pol}, this situation essentially never holds in SNe because the physical conditions differ from those in stars or illuminated nebulae or disks etc. During the photospheric phase, the SN radiation escapes from somewhere within the ejecta (rather than the ejecta base because the ejecta store the energy to be released over a large volume) and over a length scale that depends on wavelength, time, ionization etc. Further, prior to the recombination epoch in Type II SNe, the ionization is not zero anywhere in the ejecta so that scattering occurs even above the photosphere -- this effect is greater if \nifs\ is present at larger ejecta velocities  (Fig.~\ref{fig_1d_ejecta}). Thus the radiation is emitted and scattered over a sizable length scale. In contrast, at the recombination epoch, the formation of a steep recombination front implies that the flux changes mostly, and abruptly across the front. At late times, when the ejecta turn nebular, the power source covers an extended region set by the distribution of \nifs\ --- this spatial region is further extended when the $\gamma$-ray mean free path increases and causes the nonlocal deposition of decay power. These behaviors are illustrated for models e1ni1 and e1ni1b2 in Fig.~\ref{fig_flux_vs_vel} and show that even in 1D the radiation from the SN ejecta does not arise from a localized, central point source, nor from a well defined, narrow, layer such as the photosphere. These fundamental properties are important to consider when interpreting SN polarization.

 \begin{table}[h]
\caption{Summary of radiative transfer simulations in both 1D (upper part) and 2D (lower part).
\label{tab_set}
}
\begin{center}
\def\arraystretch{1.0}
\begin{tabular}{l@{\hspace{2mm}}l@{\hspace{2mm}}
c@{\hspace{2mm}}c@{\hspace{2mm}}c@{\hspace{2mm}}}
\hline
\multicolumn{5}{c}{1D simulations with \cmfgen} \\
\hline
Dim.      & Model & $E_{\rm kin}$    &     $M(^{56}$Ni)$_{\rm core}$      & $M(^{56}$Ni)$_{\rm shell}$     \\
                          &                & [10$^{50}$\,erg]                   &    [\msun] &    [\msun]   \\
\hline
1D                    & e1ni1       & 2.0         & 0.009   & 0.0   \\
1D                    & e1ni2       & 2.0         & 0.050   & 0.0   \\
1D                    & e1ni1b1   & 2.0         & 0.009   & 0.02   \\
1D                    & e1ni1b2   & 2.0         & 0.009   & 0.05   \\
\hline
1D                    & e2ni1       & 4.0         & 0.009   & 0.0   \\
1D                    & e2ni2       & 4.0         & 0.050   & 0.0   \\
1D                    & e2ni1b1   & 4.0         & 0.009   & 0.02   \\
1D                    & e2ni1b2   & 4.0         & 0.009   & 0.05   \\
\hline
\end{tabular}
\begin{tabular}{l@{\hspace{2mm}}l@{\hspace{4mm}}c@{\hspace{4mm}}
c@{\hspace{4mm}}c@{\hspace{4mm}}c@{\hspace{4mm}}}
\hline
\multicolumn{6}{c}{2D simulations with \longpol} \\
\hline
Dim.      & Model &   \multicolumn{4}{c}{Geometrical setup}  \\
2D                     & Y/e1ni1   & \multicolumn{4}{c}{Model Y for $\mid\theta\mid$\,$\lesssim$\,28.13$^\circ$,}  \\
                          &                & \multicolumn{4}{c}{e1ni1 elsewhere}  \\
\hline
Dim.    &           Model    Y        &   \multicolumn{3}{c}{$P_{\rm cont,max}$ [\%]}  & Sign Flip?\\
                        &                                  &          50\,d     & Max  & 300\,d  &     \\
\hline
2D                   &          e1ni2    & 0.10 & 1.03 &  0.93 & Yes \\
2D                   &        e1ni1b1  & 0.05 & 1.72 &  0.81 & Weak \\
2D                   &        e1ni1b2  & 0.09 & 1.83 &  1.24 & No \\
2D                   &          e2ni1    & 0.57 & 1.90 &  0.21 & Marginal \\
2D                   &          e2ni2    & 0.59 & 1.86 &  1.41 & Marginal \\
2D                   &        e2ni1b1  & 0.98 & 4.06 &  0.80 & Marginal \\
2D                   &        e2ni1b2  & 1.19 & 2.71 &  1.11 & No \\
\hline
\end{tabular}
\end{center}
Notes: The first column gives the dimensionality of the simulation. The 1D, spherically-symmetric, non-LTE time-dependent simulations with \cmfgen\ are used as initial conditions for the 2D, axially-symmetric, polarized radiative transfer simulations with \longpol. The second column gives the model name. The following columns provide some characteristics of each simulation. For the 1D simulations, we give the ejecta kinetic energy and the total mass of \nifs\ (it may be present in the core and in an external shell; see Fig.~\ref{fig_1d_ejecta}). For the 2D simulations, we give the continuum polarization at 50 and 300\,d, as well as its maximum value attained. Also indicated is whether the continuum polarization exhibits a sign flip during its evolution from 15 to 300\,d. The evolution of the quantity $-Q_{\rm cont}$, defined as $-100 F_Q/F_I$, is shown in Fig.~\ref{fig_qcont} -- what is shown is the average of that quantity over the spectral region from 6900 to 7200\,\AA. The full description of these quantities is given in the Appendix~\ref{sect_nomenclature}. All 2D simulations adopt mirror symmetry with respect to the equatorial plane.
\end{table}

\subsection{Axially-symmetric calculations with \longpol}
\label{sect_2d}

\begin{figure*}[t]
\begin{center}
\epsfig{file=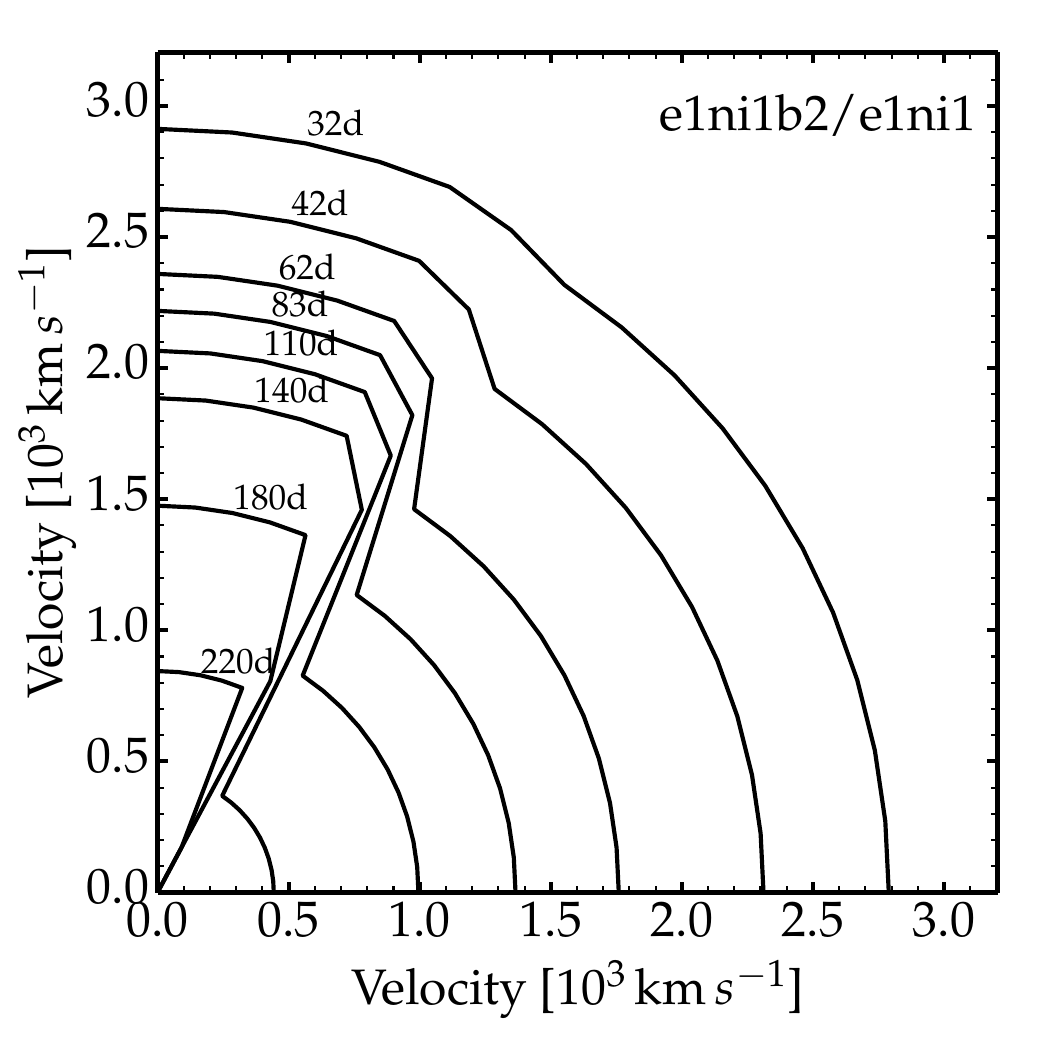, width=6.18cm}
\epsfig{file=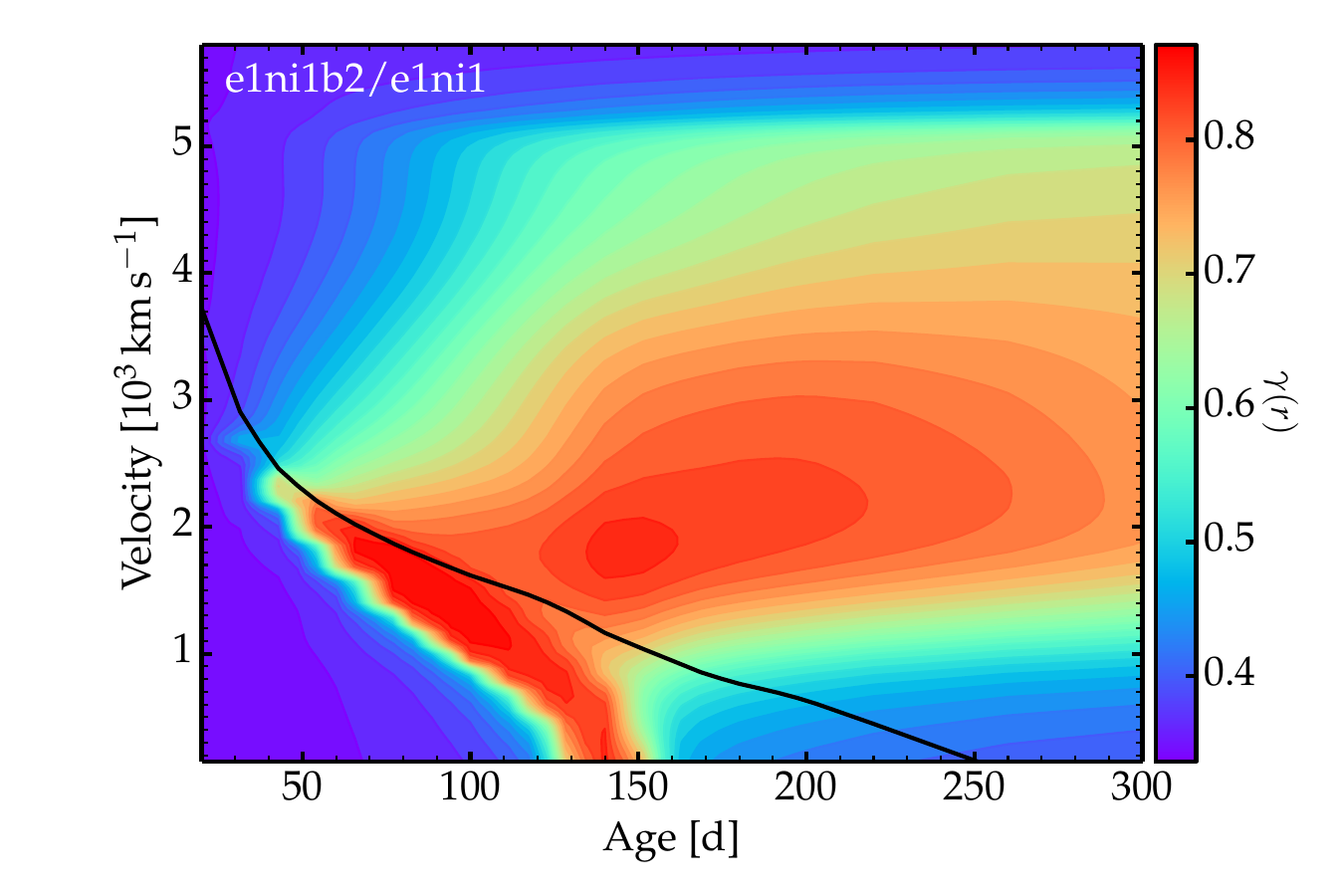, width=9.2cm}
\caption{Left: Illustration of the evolving morphology of the electron-scattering photosphere (obtained through a radial integration of the electron-scattering optical depth) in the 2D axisymmetric model e1ni1b2/e1ni1. The axis of symmetry lies in the vertical direction. Only one octant is shown since our 2D simulations adopt mirror symmetry with respect to the equatorial plane. Right: Corresponding evolution of the shape factor $\gamma(r)$, shown as a colormap, between 32 and 300\,d, and as function of velocity (because of homologous expansion, the radius is equal to the velocity multiplied by the SN age -- radius and velocity are therefore analogous quantities). Spherical symmetry corresponds to $\gamma(r)=$\,1/3. The black line gives the representative, angle-averaged location of the photosphere at each epoch.
\label{fig_2d_phot}
}
\end{center}
\end{figure*}

The 1D \cmfgen\ models presented in the preceding section are mapped onto 2D axially-symmetric ejecta (i.e., meridional slices), following the same procedure (including grid setup, radial and angular resolution etc) as described for the modeling of Type II-P SN\,2012aw in Section~3.2.3 of \citet{dessart_12aw_21}. Because the low-energy low-\nifs\ mass model e1ni1 is suitable for SN\,2008bk (see previous section, Fig.~\ref{fig_spec_1d_08bk}, and \citealt{lisakov_08bk_17}), this model represents most of the 3D ejecta volume in our setup, while the source of the asymmetry represented by one of the other seven models occupies a small solid angle, chosen in all cases here to lie within $\beta_{c}$\,$\approx$\,28.13$^\circ$ of the axis of symmetry. Specifically, we use model e1ni1 to cover polar angles between 33.75$^\circ$ and 90$^\circ$ and one of the other seven models to cover between zero and 22.50$^\circ$. Linear interpolation between the two models is used to define the properties of the 2D ejecta between 22.50 and 33.75$^\circ$. The quantities from the 1D \cmfgen\ simulations that are mapped onto the 2D ejecta are the electron density, as well as the total opacity and emissivity at all depths and wavelengths. Top-bottom, mirror symmetry is adopted so our 2D, axially-symmetric structures are, roughly speaking, similar to oblate or prolate ellipsoids. The left panel of Fig.~\ref{fig_2d_phot} illustrates the evolution of the 2D contour of the location of the electron-scattering photosphere (as given by an integration along radial, ejecta centered, rays) from 32 until 200\,d after explosion for the 2D model e1ni1b2/e1ni1.

For our 2D ejecta models, we adopt prolate configurations with a higher kinetic energy or higher \nifs\ mass along the polar direction. These configurations capture some of the features seen in 3D explosion simulations (e.g., \nifs\ fingers extending into the outer ejecta etc.; \citealt{gabler_3dsn_21}). In our 2D geometry, an oblate configuration for the higher energy or \nifs-rich material would correspond instead to a structure with a full 2$\pi$ lateral coherence, as in a disk or a torus, which seems more contrived.

As nature produces a much greater variety of ejecta configurations  our choice of eight models covers only a small fraction of the existing diversity. Despite this limitation, the present multiepoch polarization calculations are the first of their kind. Our choice of varying the explosion energy and the \nifs\ mass treats two of the most fundamental parameters known to vary amongst Type II-P SNe (see, for example, \citealt{pejcha_prieto_sn2p_15}). Varying the progenitor mass is unlikely to be a major factor since most Type II-P SN progenitors die with a comparable H-rich envelope mass \citep{d19_sn2p}.  Exploration on the influence of the opening angle, top-bottom symmetry, or explosion energy on the polarization were already explored at photospheric epochs in \citet{dessart_12aw_21} and at nebular epochs in \citet{dessart_pol_blob_21}.

The present study is a conceptual and controlled experiment that aims to delineate the diversity of continuum polarization that may arise from the documented initial conditions. Future work will be devoted to using physically consistent 3D explosion models as initial conditions but one should be aware that such simulations have their own limitations and few have ever been evolved past a few seconds after core collapse.

Throughout this work, we will also use the shape factor $\gamma(r)$ to characterize the magnitude of the asymmetry. We use a modified version of the shape factor introduced by \citet{Brown_McLean_77} such that the integral over space is performed outwards from the radius $r$ rather than over the full ejecta. Our shape factor $\gamma(r)$ is thus defined as
\begin{equation}
\gamma(r) = { \int_r^\infty \int_{-1}^{1} N_e(r,\mu) \mu^2 \,d\mu \,dr \over \int_r^\infty \int_{-1}^{1} N_e(r,\mu)  \,d\mu \,dr\,}\,\,\, , \label{eq_gamma_r}
\end{equation}
where $N_e(r,\mu)$ is the free-electron density at $(r,\mu)$ ($\mu$ is the cosine of the polar angle $\beta$). Spherical symmetry corresponds to $\gamma(r)=1/3$, with prolate (oblate) configuration corresponding to values between 1/3 and 1 (1/3 and 0). A value of one corresponds to a ``polar line'' and zero corresponds to an equatorial disk.

The shape factor was originally introduced by \citet{Brown_McLean_77} to quantify the continuum linear polarization from an asymmetric distribution of free electrons around a point source under optically-thin conditions. In our SN models, optically thin conditions are only met in the nebular phase, hence at times greater than 130--190\,d depending on the model (Fig.~\ref{fig_1d_lc}). As discussed in Appendix C and E of \citet{dessart_pol_blob_21}, truly optically thin conditions in the context of polarization tend to occur even later, when the total ejecta optical depth has dropped to about 0.1 or lower -- above such values, optical depth effects continue to operate and can quench the polarization. When looking at the shape factors, optically-thick regions play a subdominant role. Hence, in the right panel of Fig.~\ref{fig_2d_phot} and analogs, the region beyond the photosphere (indicated by the black line) is most relevant for the interpretation of the continuum polarization.

 The total \nifs\ mass or kinetic energy in our set of 2D, axially-symmetric ejecta models is equal to the \nifs\ mass of the polar model weighted by (1-$\cos \beta_c$) and the model used for other latitudes (i.e., always model e1ni1) weighted by  $\cos \beta_c$. Hence our set of 2D models cover ejecta kinetic energies between 2.0 and $2.2 \times 10^{50}$\,erg and \nifs\ masses between 0.011 and 0.015\,\msun.

The post-processing of 1D \cmfgen\ models with the 2D \longpol\ code is not fully consistent because the levels populations at different ejecta velocities and latitudes in the 2D model are held fixed during the 2D computation with \longpol. The asset of \longpol\ is to compute the 2D radiation field for the imposed 2D distribution of opacities and emissivities within the 2D ejecta, and in particular the 2D distribution of free electrons.

In this work, we focus exclusively on ``continuum'' polarization, and select the spectral region from 6900 to 7200\,\AA.  Results are to be compared with spectropolarimetric observations, which dominate the observational literature on SN polarization studies. They are not to be used for comparison with broad-band polarimetric observations. The region from 6900 to 7200\,\AA\ is the one commonly used in the observational literature. In practice, \longpol\ computes the polarization throughout the optical range accounting explicitly for the influence of line and continuum processes (see, e.g., \citealt{dessart_12aw_21}). The region between 6900 and 7200\,\AA\ is relatively line free and therefore is at all times weakly affected by metal line blanketing (see Fig.~\ref{fig_spec_1d_08bk}). Broadening this range from 6900 to 8200\,\AA\ (as used for example by \citealt{leonard_04dj_06}) yields better statistics but this range then covers for example the \caiidoub, which is strong at nebular times. We have tested with each choice and find that it introduces a modest quantitative offset in the continuum polarization versus time (see  Fig.~\ref{fig_cont_range}). Furthermore, the scattered H$\alpha$ line flux could be at most redshifted by twice the maximum velocity of about 5000\,\kms\ in the present models so that leads to a redshifted flux from H$\alpha$ out to a maximum wavelength of about 6800\,\AA. Hence, the scattered H$\alpha$ photons cannot affect the measured polarization beyond 6900\,\AA.

\begin{figure*}[t]
\begin{center}
\epsfig{file=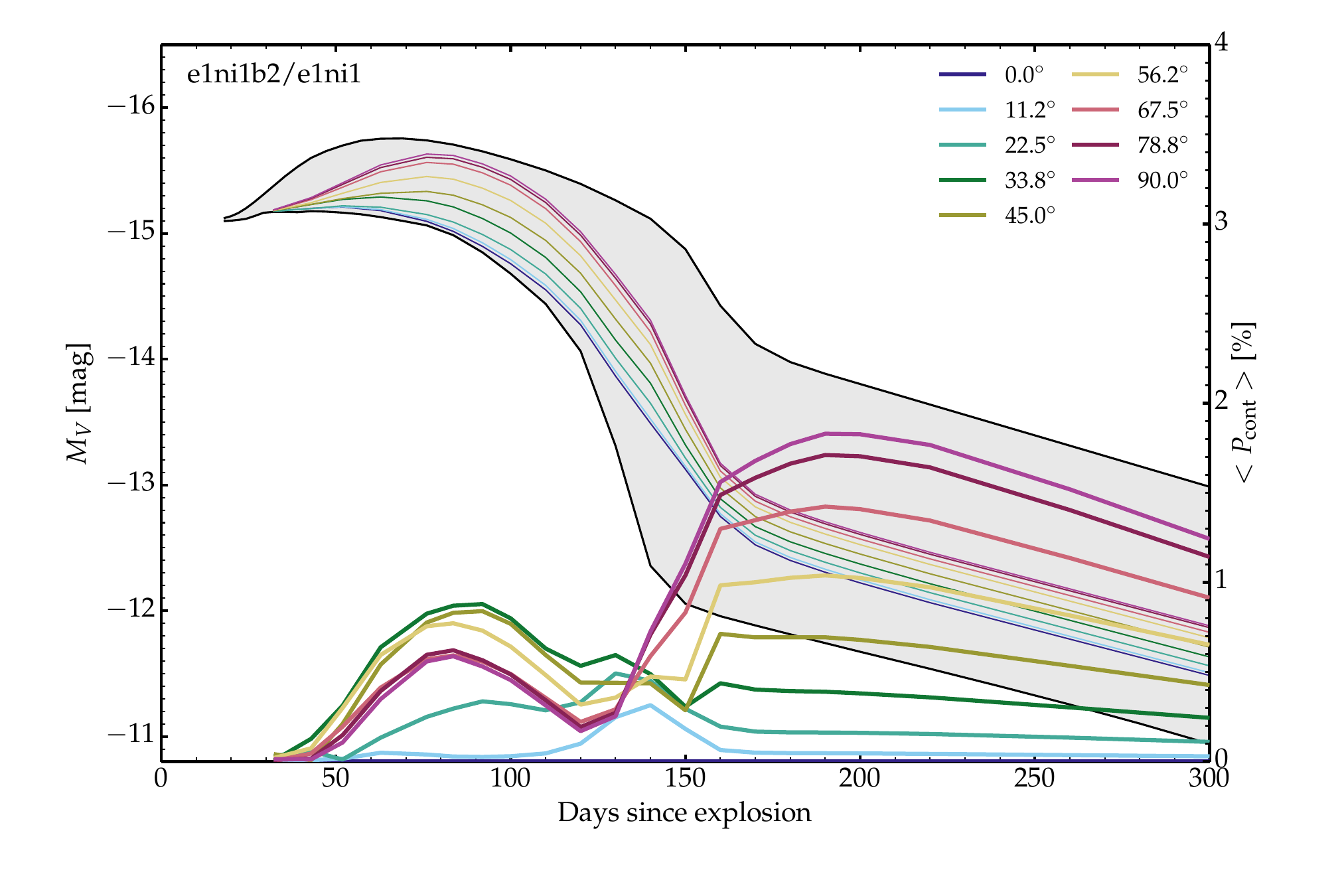, width=17cm}
\vspace{-0.8cm}
\caption{Results for the $V$-band light curve and the evolution of the continuum polarization for the 2D model e1ni1b2/e1ni1. We show the evolution of the $V$-band magnitudes (thin lines; yaxis at left) and continuum polarization (thick line; yaxis at right) from 20 to 300\,d. The 2D model has the properties of the 1D model e1ni1b2  for polar angles within about 28$^\circ$ of the axis and those of the 1D model e1ni1 elsewhere. The different colors refer to different viewing angles relative to the axis of symmetry, covering from 0$^\circ$ (pole-on) to 90$^\circ$ (edge-on) -- see legend at top-right. The grey area is bounded by the \cmfgen\ light curves computed for 1D models e1ni1 (bottom black curve) and e1ni1b2 (top black curve). The continuum polarization corresponds to polarization in the relatively line-free region between 6900 and 7200\,\AA. \longpol\ finds  a null polarization at all epochs for a zero degree inclination, as expected.
\label{fig_e1ni1b2_e1ni1}
} 
\end{center}
\end{figure*}

\section{Results for a representative 2D model}
\label{sect_pcont_ref}

Figure~\ref{fig_e1ni1b2_e1ni1} shows results from \longpol\ for the evolution of the $V$-band magnitude\footnote{We illustrate the $V$-band magnitude, rather than a pure continuum magnitude, since the latter is not routinely observed.} and the continuum polarization from 20 to 300\,d after explosion and for different viewing angles for model e1ni1b2/e1ni1\footnote{The naming convention used here and for all 2D models is defined in Table~\ref{tab_set}. Specifically, the model name lists the model inserted into the polar regions within $\approx$\,28.13$^\circ$ first (here e1ni1b2) followed by the base model employed for the rest of the ejecta (here, and in all cases, e1ni1).}(corresponding to the 2D ejecta shown in Fig.~\ref{fig_2d_phot}).  For different viewing angles, the $V$-band light curve lies between (and typically well away from) that of the 1D model e1ni1 (lower-black curve) and that of 1D model e1ni1b2 (upper black curve). For a pole-on view, the radial extent of the electron-scattering photosphere, which is essentially at the H-recombination temperature, is unchanged relative to model e1ni1 so a distant observer records essentially the same brightness. For an equator-on view, the radial extent of the electron-scattering photosphere is greater along the pole, increasing the size of the radiating surface relative to that seen from the pole-on view. For example, at 84\,d, the photospheric radius is 60\,\% bigger along the pole and the equator-on view of the radiating surface is about 50\,\% larger, corresponding to a 0.5\,mag offset. This contrast in the size of the radiating surface with viewing angle causes the bump in $V$-band brightness in the second-half of the photospheric phase for viewing angles closer to 90$^\circ$. This bump also arises because in our low energy model, the plateau brightness (in the absence of \nifs) is quite low so a large \nifs\ mass as employed in model e1ni1b2 can significantly impact the otherwise faint plateau brightness. In a standard explosion model, this photometric boost from \nifs\ decay heating would have had a weaker impact because low-energy explosions produce less \nifs.

Model e1ni1b2/e1ni1 has more \nifs\ than model e1ni1 but less than model e1ni1b2 so the ejecta transition to the nebular phase at a time intermediate (i.e., about 160\,d and about the same for all viewing angles) between those obtained for these two 1D models. The late-time brightness is systematically greater for lower latitudes with a 0.35\,mag offset between a pole-on and an equator-on view. The persisting dependence of the luminosity with viewing angle is surprising at nebular times and suggests that there are still optical depth effects. There would be no photometric variation with viewing-angle if the optical depth was zero but in this model $\tau_{\rm es}$ is 0.48 along the pole and 0.25 along the equator direction. There is clearly residual scattering within the ejecta because \longpol\ also predicts a nonzero polarization signal, with a strong dependence on viewing angle. This issue arises because the term optically thin is ambiguous. While it applies for any configuration having an optical depth less than 2/3, ejecta with an optical depth of 0.1 or 0.001 differ from a radiative-transfer point of view.

The continuum polarization for model e1ni1b2/e1ni1 is initially zero but rises and forms a bump at 80\,d after explosion, essentially peaking at the same time as the bump seen in $V$ band magnitude.\footnote{As discussed above for the $V$-band light curve, a smaller amount of \nifs\ would  yield a smaller polarization bump because of the reduced excess in the free-electron density -- see next section and discussion for the model e1ni1b1/e1ni1. The impact of a higher explosion energy, more typical of a standard Type II-P SN, on the polarization is less clear because in that case the mass density, the ionization, and thus the free-electron density would change, even without a high velocity \nifs\ blob. Results for that case have, however, been presented in \citet{dessart_12aw_21}.} The polarization at that time is maximum for mid-latitudes, which indicates that there are optical depth effects (for an optically-thin configuration, the polarization would be maximum for a 90$^\circ$ inclination or viewing angle). The polarization then drops to a minimum at 120\,d before rising again. For most viewing angles, the maximum polarization is reached at the onset of the nebular phase as defined by the time when the model luminosity falls on the \cofs\ decay tail phase. However, for low latitudes, the continuum polarization rises until about 200\,d, reaching a maximum of $\sim$\,1.8\% for an equator-on view. The rate at which the continuum polarization decreases at late times is much slower than $1/t^2$.  For high latitudes, the continuum polarization stays nearly constant, while along low latitudes, the decrease is close to $1/t$. This behavior arises from a variety of reasons. First, the ejecta optical depth is not far from unity so conditions are not strictly optically thin, as assumed in \citet{Brown_McLean_77}. This issue in the context of SN polarization is well known (see, for example, \citealt{hoeflich_87A_91}; \citealt{DH11_pol}; \citealt{dessart_pol_blob_21}). The ionization is not constant but the ejecta recombination (i.e. the rate of change of the ejecta ionization) at nebular times is very slow so that probably plays little role here (see Fig.~\ref{fig_tau_es_e1ni1}). More important is the extended emission from the ejecta so that both emission and scattering occur in the same volume, in contrast with the point-source approximation assumed in \citet{Brown_McLean_77}.

In this 2D model, the continuum polarization $Q_{\rm cont}$ does not change sign during the whole evolution (top-right panel in Fig.~\ref{fig_qcont}). It stays negative at all times and for all viewing angles. With our sign convention (see Appendix~\ref{sect_nomenclature}), this indicates that the electric vector is perpendicular to the axis of symmetry, as expected for the prolate 2D ejecta configuration of model e1ni1b2/e1ni1.

Further insight can be gained by inspecting the shape factor for model e1ni1b2/e1ni1 (right panel of Fig.~\ref{fig_2d_phot}). The presence of a \nifs\ blob along the polar direction boosts the electron density around 2000\,\kms\ (compare the curves for models e1ni1b2 and e1ni1 in Fig.\ref{fig_1d_ejecta}). This leads to an asphericity of the free-electron density as well as the electron-scattering photosphere (as given by a radial integration of the electron-scattering optical depth) after about 40\,d. The shape factor $\gamma(r)$ exhibits a relatively narrow peak with a maximum at the photosphere at about 80\,d, which is the time of the first polarization maximum --- $\gamma(r)$ already reaches 0.83 at an age of 63\,d but $\gamma(r)$ is relatively low above the photosphere. However, at late times, $\gamma(r)$ stays high at the photosphere and is also high (i.e., about 0.7) throughout the envelope (i.e., $>$\,1000\,\kms), with a maximum around 200\,d. These variations reflect roughly the evolution of the continuum polarization shown in Fig.~\ref{fig_e1ni1b2_e1ni1}. The analogy is only approximate because of optical depth effects, which are strong during the photospheric phase and also persist until late times. 

Finally, we see that the shape factor integrated all the way to the ejecta base is much closer to 1/3. The inner ejecta in model e1ni1b2/e1bni1 is nearly spherical and because the electron density is much larger in those layers, it dominates in the integrals of Eq.~\ref{eq_gamma_r}. The polarization signature, including the jump at the onset of the tail phase is thus clearly not tied to the asymmetry of the core, which is small, but by the asymmetry of the outer H-rich ejecta layers, even though the electron density is weaker in those regions. This confirms the previous findings of  \citet{DH11_pol} and \citet{dessart_12aw_21}.

Having discussed in detail the results for a representative case, we discuss in the next section the results for the rest of the sample of 2D models calculated with \longpol. The contours illustrating the morphology of the 2D electron-scattering photospheres are plotted in the appendix in Fig.~\ref{fig_2d_phot_rest}.

\begin{figure*}[h]
\epsfig{file=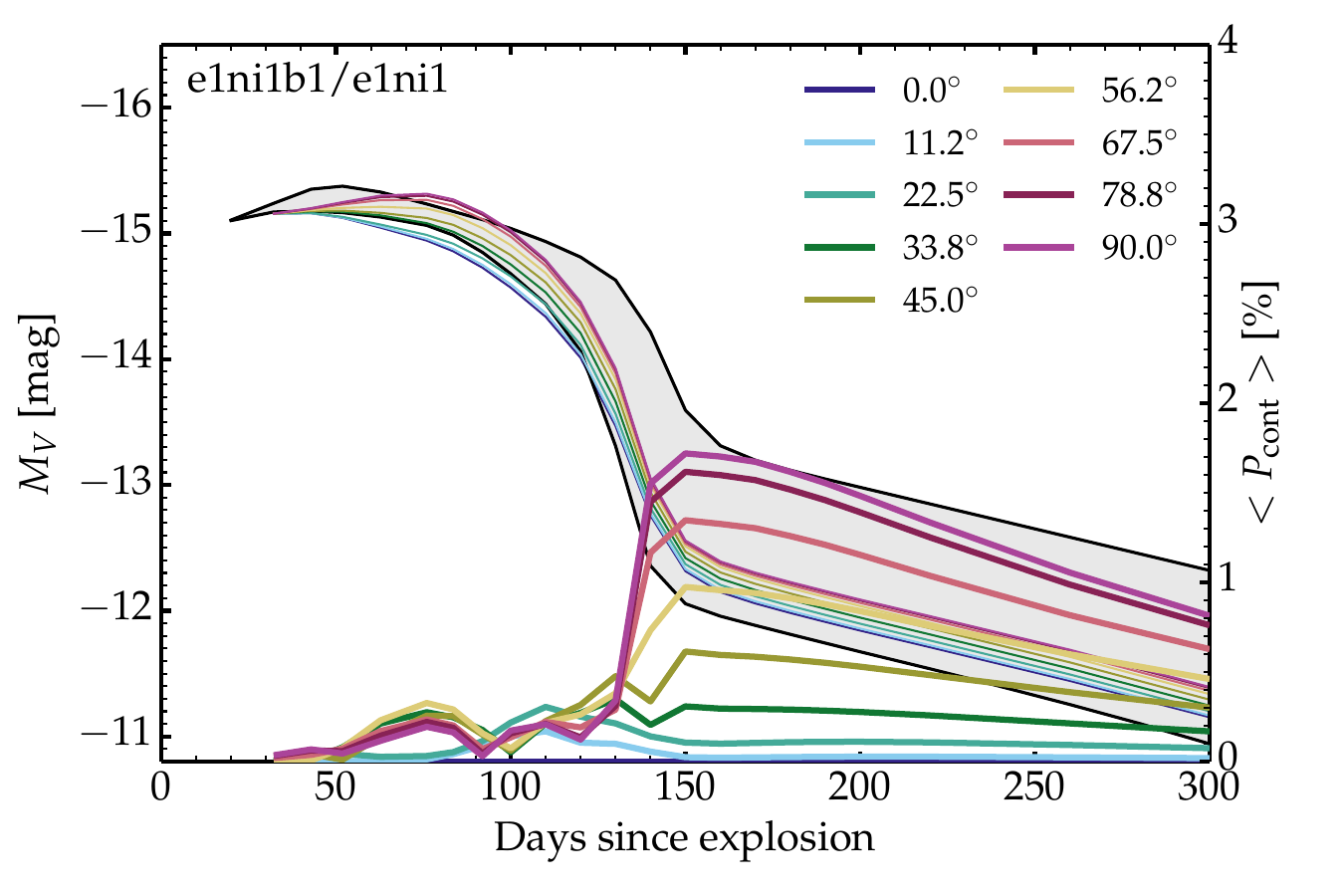, width=9.2cm}
\epsfig{file=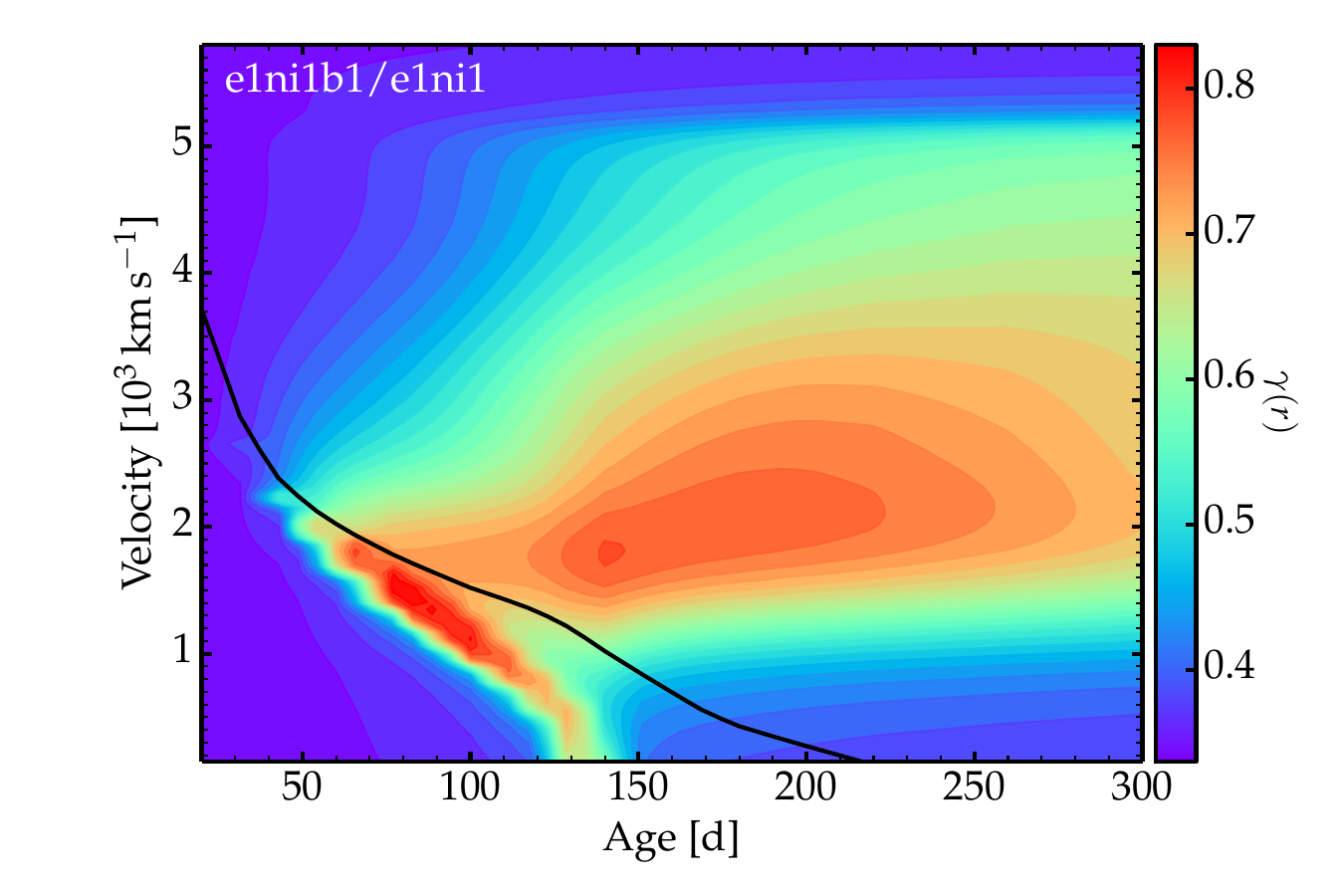, width=9.2cm}
\caption{Same as Fig.~\ref{fig_e1ni1b2_e1ni1} and Fig.~\ref{fig_2d_phot} (right panel) but showing the results for model e1ni1b1/e1ni1. [See text for discussion.]
\label{fig_e1ni1b1_e1ni1}
}
\end{figure*}

\section{The evolution diversity of continuum polarization}
\label{sect_pcont_all}

Figure~\ref{fig_e1ni1b1_e1ni1} is an analog of Fig.~\ref{fig_e1ni1b2_e1ni1} but for model e1ni1b1/e1ni1. This 2D model exhibits qualitatively similar but quantitatively different results from those for model  e1ni1b2/e1ni1 because of the slight differences in ejecta properties. The \nifs\ blob at large velocity contains less \nifs\ originally than in model e1ni1b2, so the boost to the electron density is weaker (Fig.~\ref{fig_1d_ejecta}) and the shape factor is shifted to lower values closer to 1/3 (right panel of Fig.~\ref{fig_e1ni1b1_e1ni1}). The $V$-band bump during the photospheric phase is weaker, and so is the associated bump in continuum polarization. Optical depth effects are weaker at the end of the photospheric phase and we see that the continuum polarization rises sharply for inclinations at low latitudes. The decrease during the nebular phase is also steeper. The polarization curve for the viewing angle 56.2$^\circ$ agrees roughly with that observed for SN\,2004dj \citep{leonard_04dj_06} --- although perhaps accidental, this inclination is comparable to that inferred by \citet{chugai_04dj_06} for SN\,2004dj.

\begin{figure*}
\epsfig{file=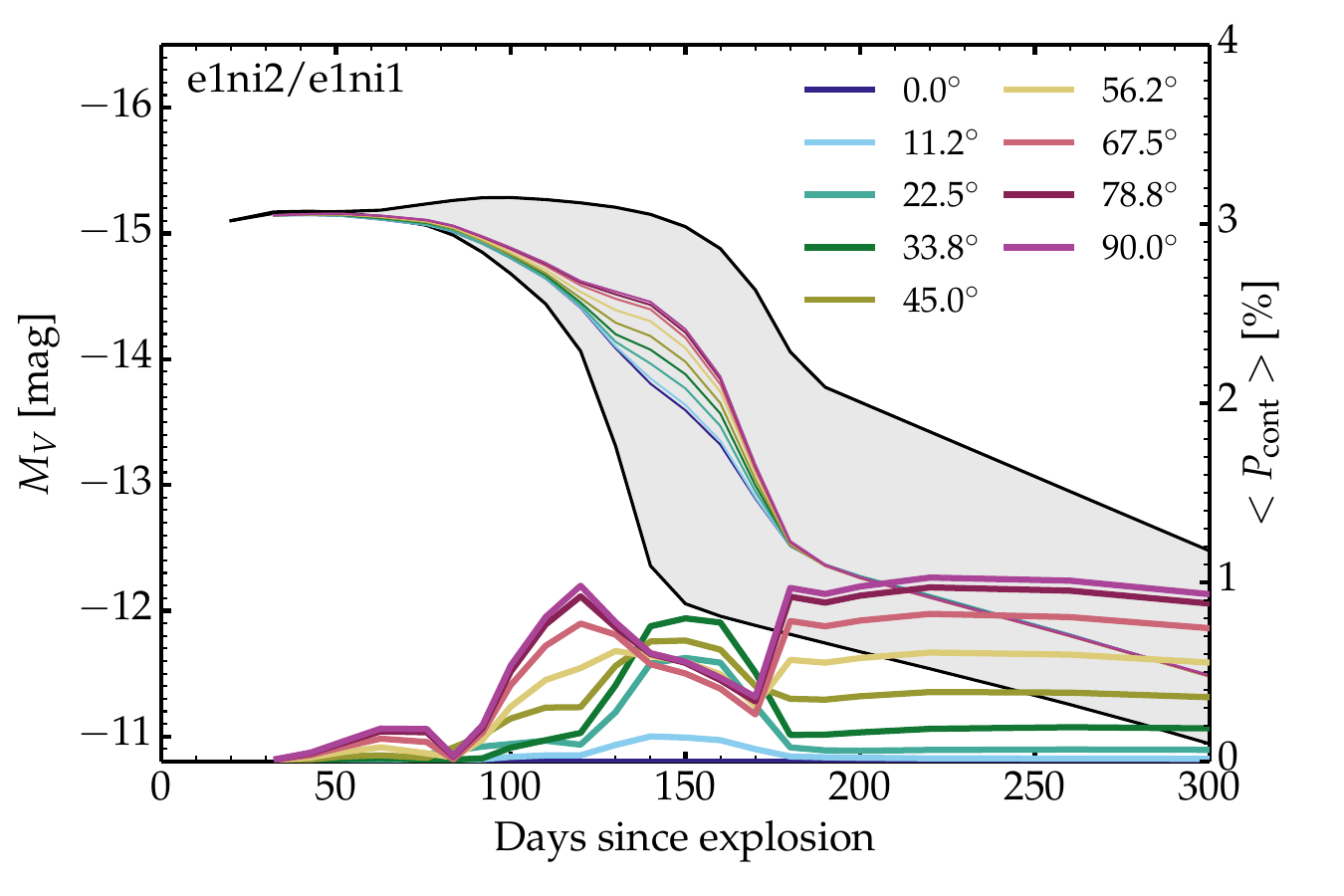, width=9.2cm}
\epsfig{file=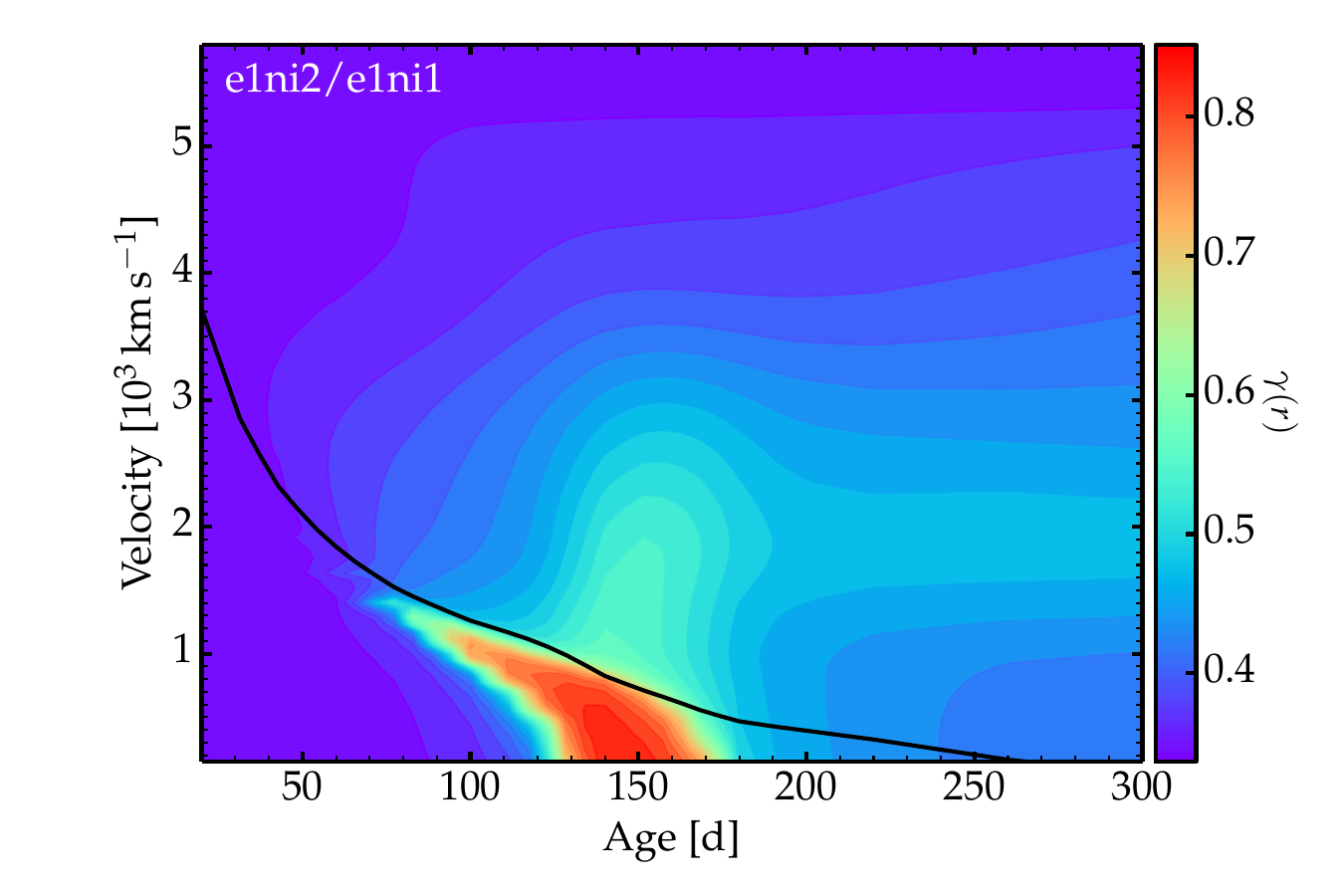, width=9.2cm}
\caption{Same as Fig.~\ref{fig_e1ni1b2_e1ni1} and Fig.~\ref{fig_2d_phot} (right panel)  but showing the results for model e1ni2/e1ni1. [See text for discussion.]
\label{fig_e1ni2_e1ni1}
}
\end{figure*}

Figure~\ref{fig_e1ni2_e1ni1} shows the results for the 2D model e1ni2/e1ni1. In model e1ni2, the enhanced \nifs\ mass is limited to the inner ejecta (Fig.~\ref{fig_1d_ejecta}), which causes an increase in the radius of the electron scattering photosphere near the end of the photospheric phase (see top-right panel of Fig.~\ref{fig_2d_phot_rest}). The $V$ band light curve is essentially independent of viewing angle except near the end of the photospheric phase around 150\,d. The shape factor in the H-rich outer layers of the ejecta is systematically smaller than in models e1ni1b1/e1ni1 or e1ni1b2/e1ni1 and we see that the maximum polarization never exceeds 1\%. This indicates that an asymmetry confined to the inner ejecta produces a lower maximum continuum polarization than configurations where the asymmetry resides within the faster-moving H-rich ejecta (as in models e1ni1b2/e1ni1 or e1ni1b1/e1ni1; see Figs.~\ref{fig_e1ni1b2_e1ni1}--\ref{fig_e1ni1b1_e1ni1}).

Optical-depth effects are important in at least two ways. First, there is a strong viewing-angle dependence of the continuum polarization, as can be seen by the opposite behavior of the polarization from 100 to 180\,d for low and high latitudes in Fig.~\ref{fig_e1ni2_e1ni1}. This is caused by a sign flip for certain directions, as illustrated by the $<-Q_{\rm cont}>$ evolution shown in the third panel from top in Fig.~\ref{fig_qcont}. Because the configuration is optically thick, the residual polarization is influenced by both the asymmetric distribution of scatterers and the flux on the plane of the sky -- their distribution on the plane of the sky is opposite (i.e., if one appears oblate, the other appears prolate) and cancellation effects are significant. Second, the continuum polarization stays constant during the nebular phase for all viewing angles. This feature arises because the innermost ejecta regions along the pole, originally rich in \nifs, stay hotter, more ionized, and optically-thick until late times (the top-right panel of Fig.~\ref{fig_2d_phot_rest} indicates that the total radial electron-scattering optical depth along the poles is still greater than $2/3$ at 220\,d). This polarization evolution is reminiscent of that observed in SN\,2008bk \citep{leonard_08bk_12}.

\begin{figure*}
\epsfig{file=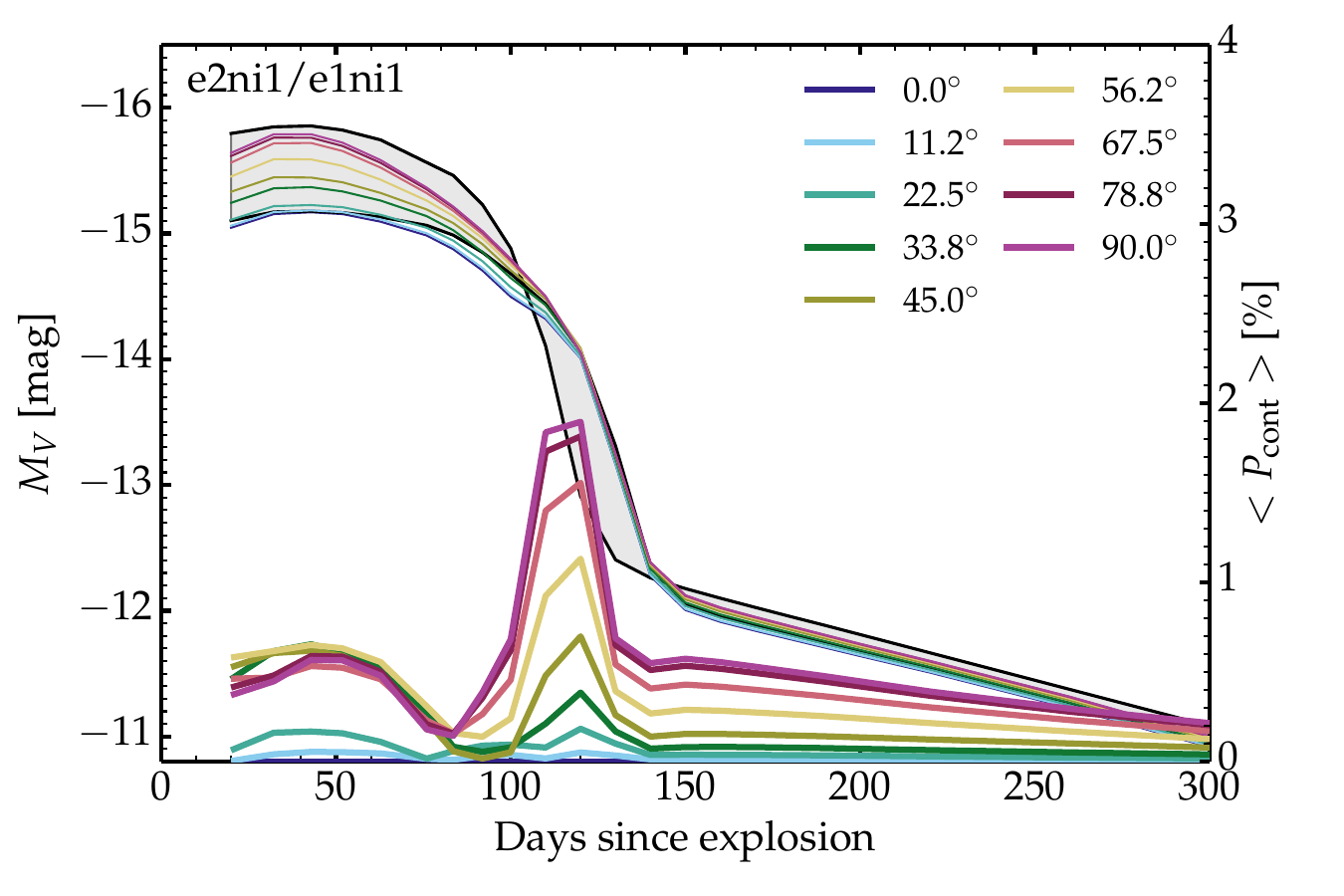, width=9.2cm}
\epsfig{file=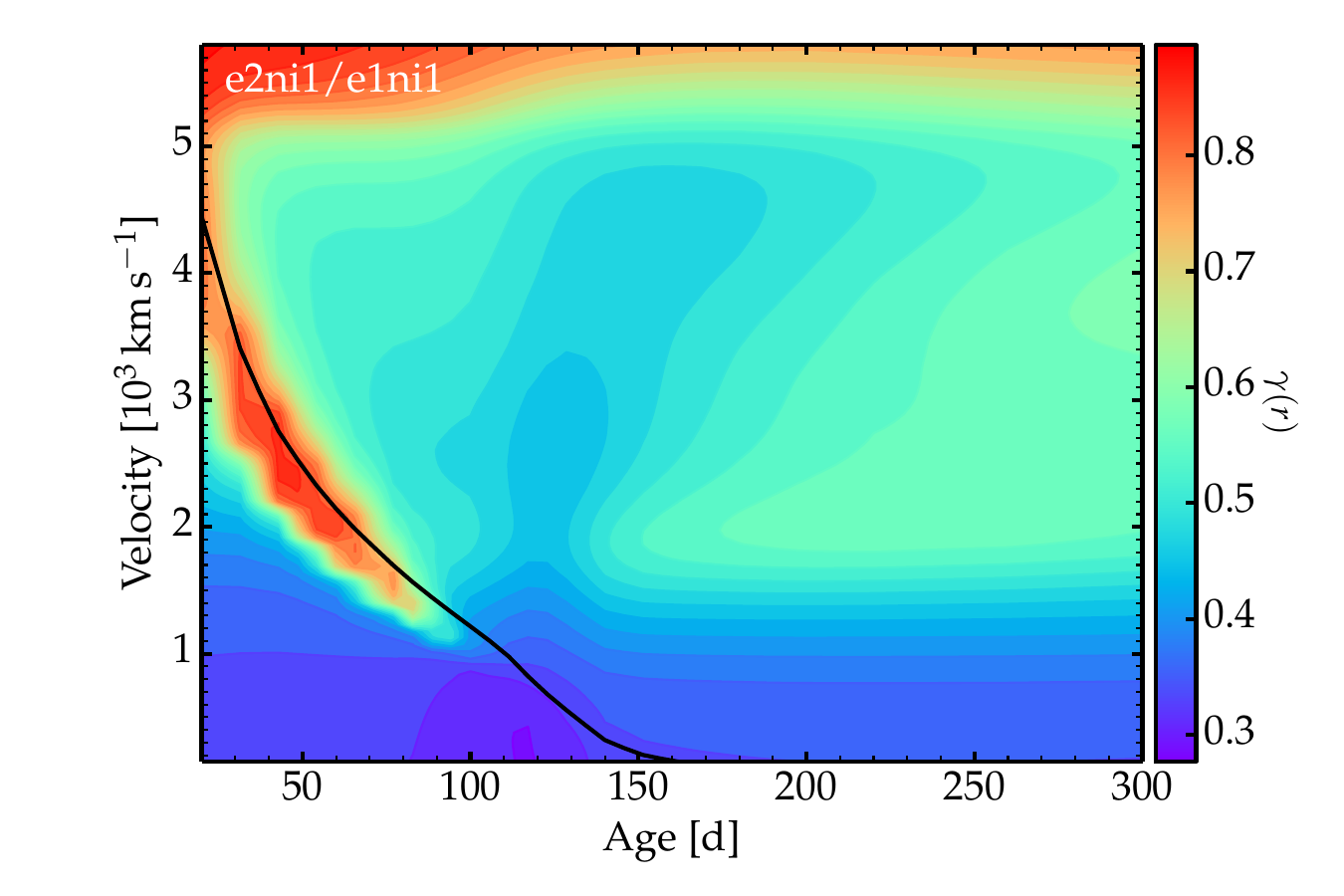, width=9.2cm}
\caption{Same as Fig.~\ref{fig_e1ni1b2_e1ni1} and Fig.~\ref{fig_2d_phot} (right panel) but showing the results for model e2ni1/e1ni1. [See text for discussion.]
\label{fig_e2ni1_e1ni1}
}
\end{figure*}

The remaining 2D configurations in our sample involve a combination of model e1ni1 with higher energy explosions. Figure~\ref{fig_e2ni1_e1ni1} shows the results for the 2D model e2ni1/e1ni1. Model e2ni1 has more mass at high velocity and less at low velocity relative to model e1ni1 so the 2D model e2ni1/e1ni1 evolves from a prolate morphology up until 110\,d to an oblate morphology thereafter (see middle-left panel of Fig.~\ref{fig_2d_phot_rest}). The higher explosion energy along the poles yields an enhanced brightness during the photospheric phase for viewing angles along low latitudes. The light curve is the same for all angles at late times and comparable to model e1ni1. The continuum polarization is nonzero throughout the photospheric phase because of the enhanced density along the poles, which causes the shape factor to reach a maximum value of 0.9 at 6000\,\kms. Because the photosphere recedes faster along the pole, the asymmetry of the 2D photosphere is continuously reduced as time passes, until it becomes  essentially spherical at 110\,d. The radial variation of the optical depth is very different both above and below the photosphere for polar and equatorial directions, as illustrated by the shape factor (right panel of Fig.~\ref{fig_e2ni1_e1ni1}), which drops below 1/3 around 110\,d at the photosphere; at both early and late times, the shape factor is always greater than $1/3$ beyond 1000\,\kms. The peak of polarization occurs at 110\,d and reaches 1.9\% for an equator-on view, even though at this time the 2D photosphere is spherical -- the ejecta regions above that spherical photosphere have a $\gamma(r)$ of 0.4--0.5 and are the cause of the residual polarization.  The shape of the photosphere is therefore not a reliable diagnostic for interpreting the polarization. When all directions turn optically thin, the polarization drops abruptly and follows a slow decrease past 150\,d. This well defined narrow maximum in continuum polarization at the onset of the tail phase is qualitatively reminiscent of what was observed in SN\,2008bk.

\begin{figure*}
\epsfig{file=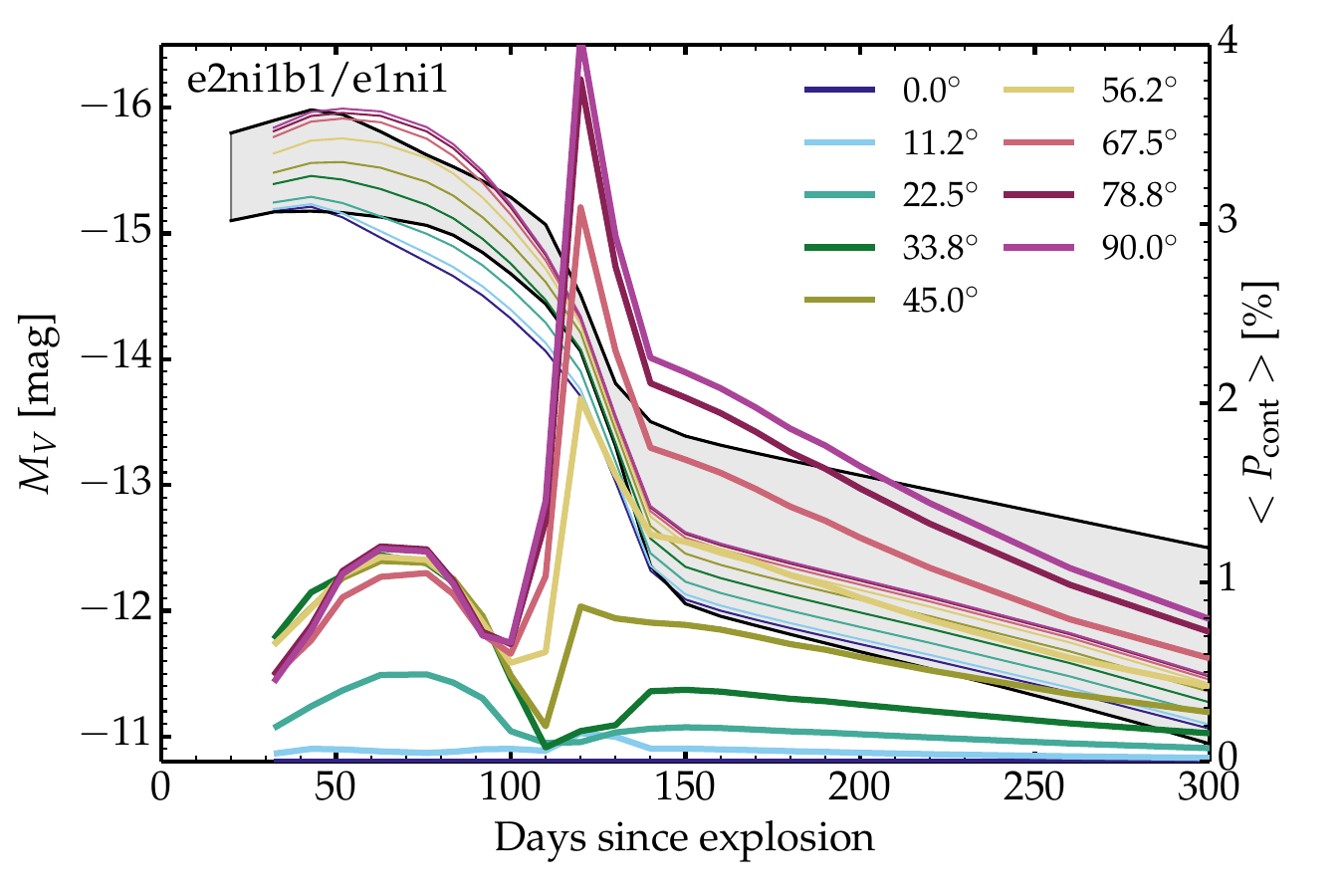, width=9.2cm}
\epsfig{file=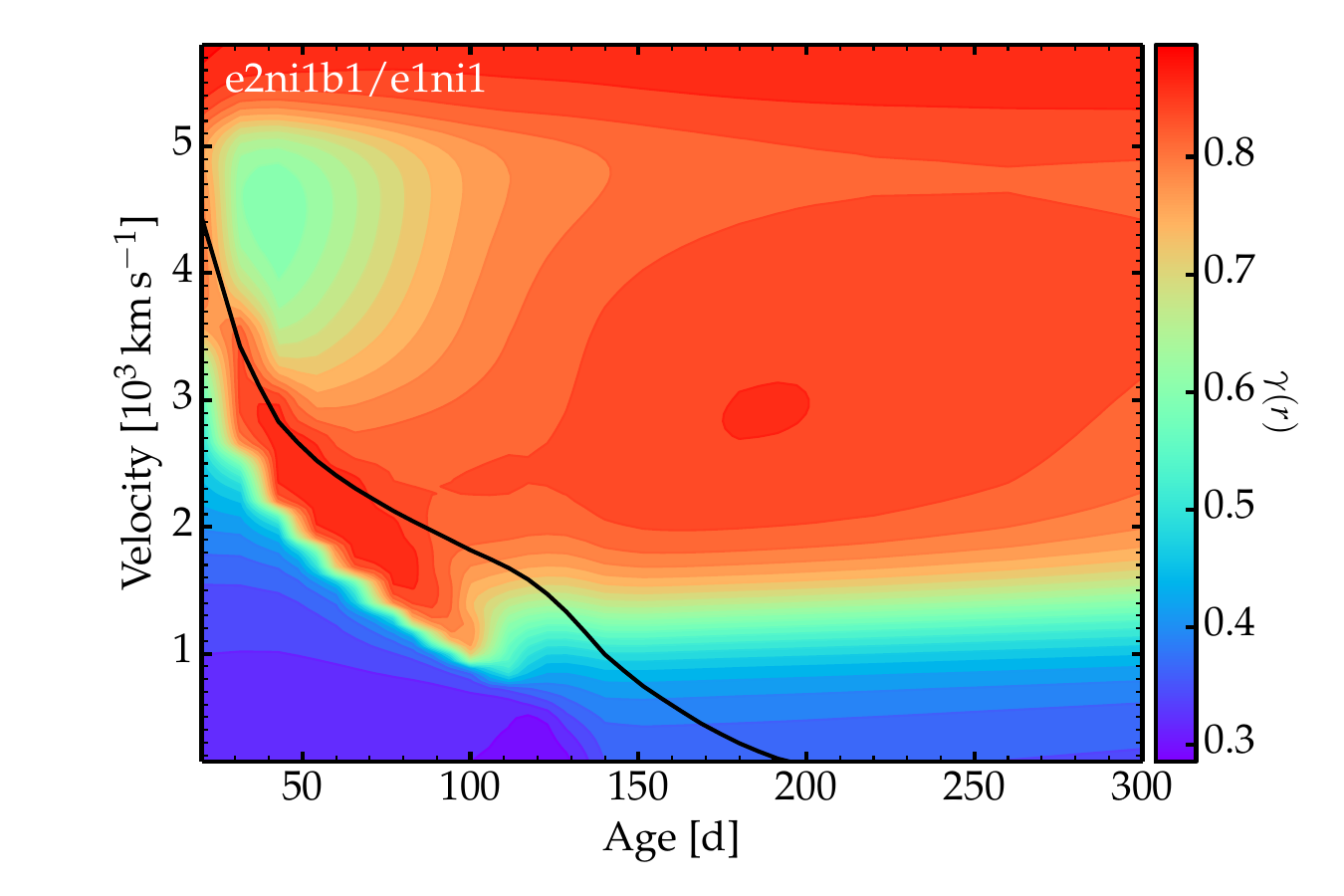, width=9.2cm}
\caption{Same as Fig.~\ref{fig_e1ni1b2_e1ni1} and Fig.~\ref{fig_2d_phot} (right panel) but showing the results for model e2ni1b1/e1ni1. [See text for discussion.]
\label{fig_e2ni1b1_e1ni1}
}
\end{figure*}

\begin{figure*}
\epsfig{file=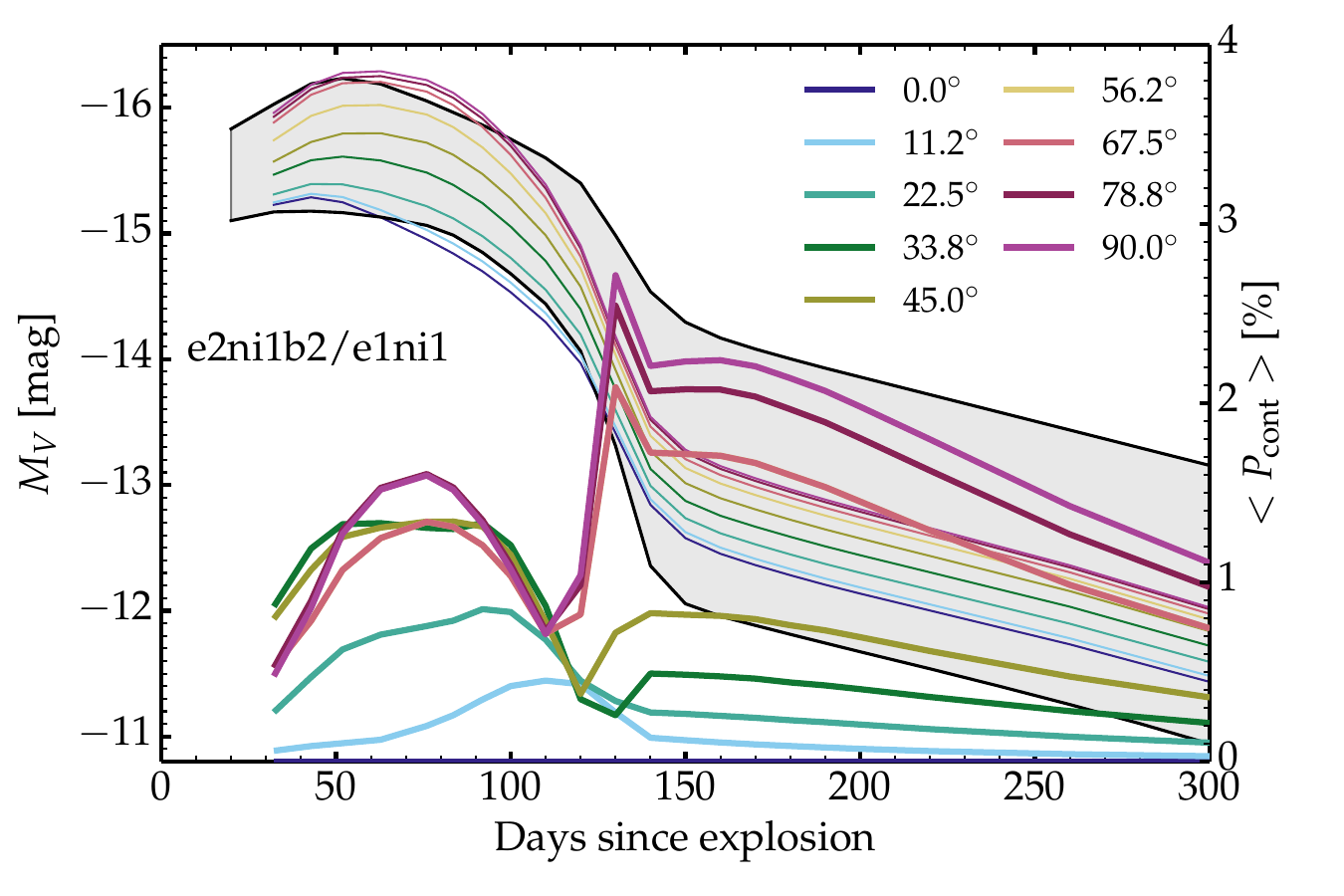, width=9.2cm}
\epsfig{file=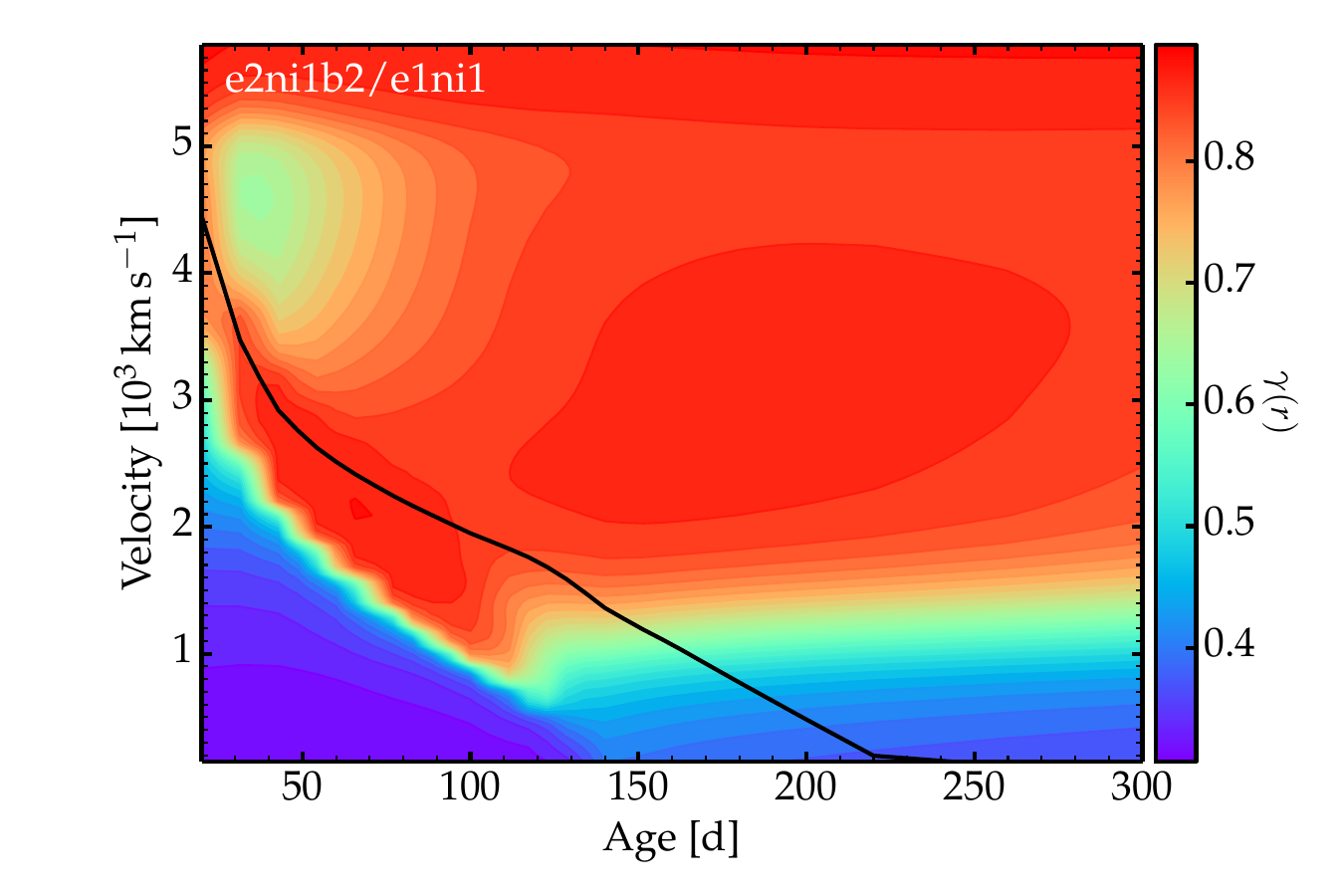, width=9.2cm}
\caption{Same as Fig.~\ref{fig_e1ni1b2_e1ni1} and Fig.~\ref{fig_2d_phot} (right panel)  but showing the results for model e2ni1b2/e1ni1. [See text for discussion.]
\label{fig_e2ni1b2_e1ni1}
}
\end{figure*}

Figures~\ref{fig_e2ni1b1_e1ni1} and \ref{fig_e2ni1b2_e1ni1} show the results for models e2ni1b1/e1ni1 and e2ni1b2/e1ni1, which are analogous. Because the higher energy explosion model also has more \nifs\ than the model e1ni1, the 2D ejecta retain a prolate morphology at all epochs. The larger \nifs\ mass boosts the ionization of the higher energy model and compensates for the faster expansion so models e2ni1b1 and e2ni1b2 stay optically thick for longer than model e1ni1 (Fig.~\ref{fig_2d_phot_rest}). This is indirectly seen in the evolution of the $V$-band light curve. Compared to model e2ni1/e1ni1, the presence of \nifs\ at high velocity maintains the shape factor near a value of 0.9 throughout the H-rich ejecta layers at $>$\,130\,d. In model e2ni1b1/e1ni1, a remarkable peak polarization of 4\% is reached at 130\,d followed by a rapid decrease close to the expected $1/t^2$ dependence. In model e2ni1b1/e1ni1, the peak is smaller at 2.7\%, which is still very large, and the decrease at nebular times is slower, most likely because of the residual optical depth of the ejecta. In both models, the polarization is still on the order of 1\% along equator-on views at 300\,d, which is comparable to the polarization maximum reached for a short time during the photospheric phase.

\begin{figure*}
\epsfig{file=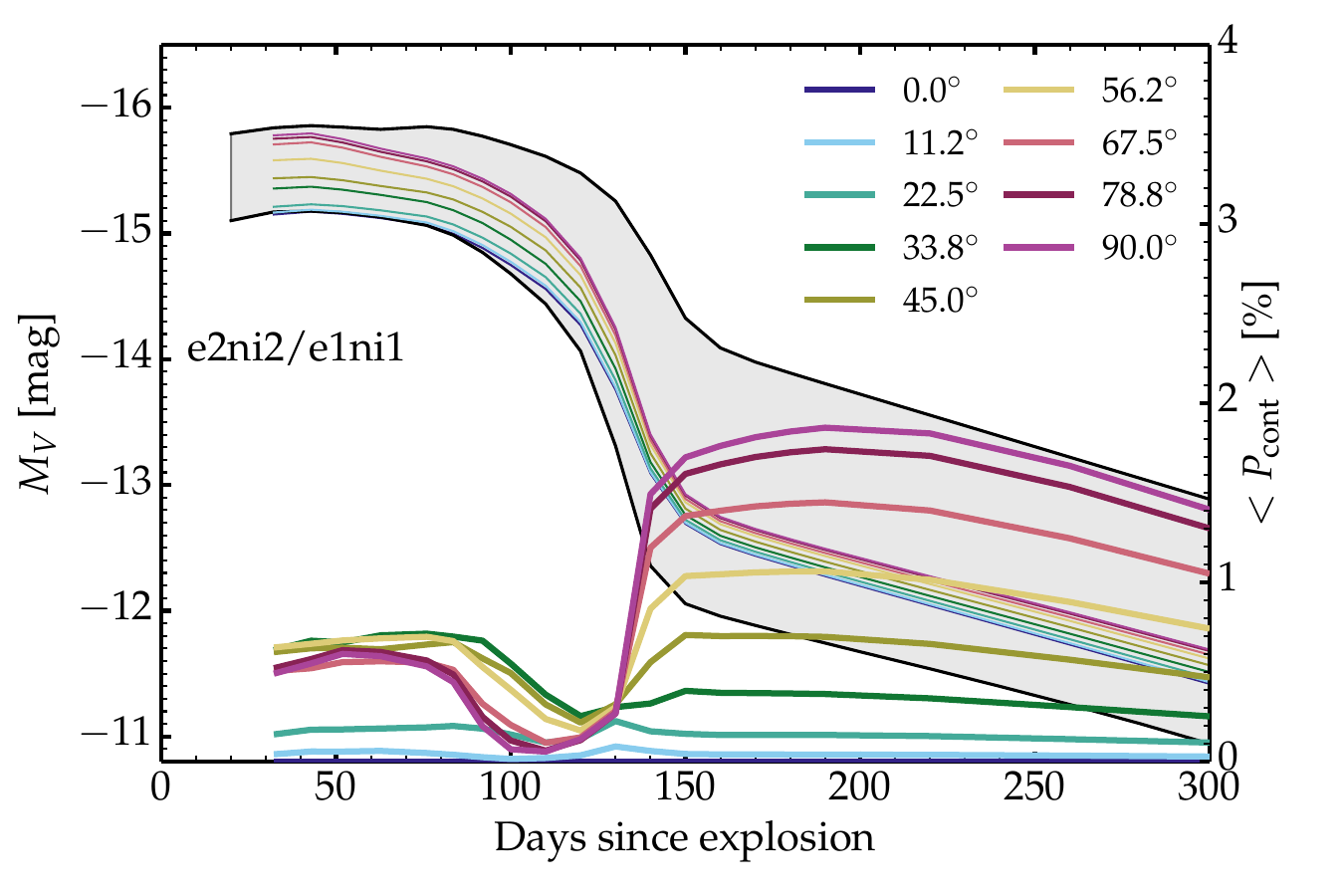, width=9.2cm}
\epsfig{file=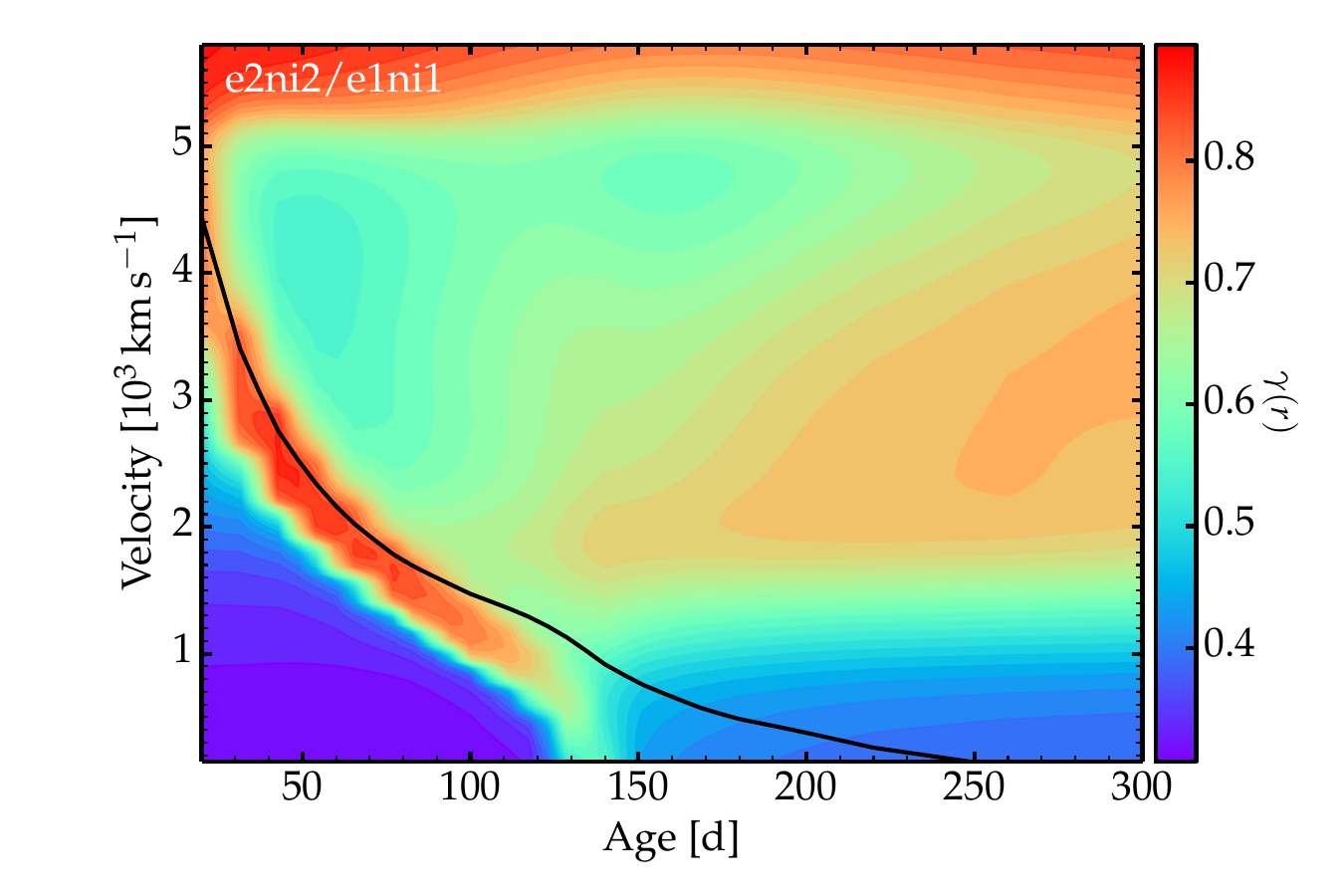, width=9.2cm}
\caption{Same as Fig.~\ref{fig_e1ni1b2_e1ni1} and Fig.~\ref{fig_2d_phot} (right panel) but showing the results for model e2ni2/e1ni1. [See text for discussion.]
\label{fig_e2ni2_e1ni1}}
\end{figure*}

Figure~\ref{fig_e2ni2_e1ni1} shows the results for the 2D ejecta model e2ni2/e1ni1. Model e2ni2 has greater kinetic energy and \nifs\ mass than model e1ni1, but the \nifs\ enhanced abundance is confined to the inner ejecta layers. While this configuration is similar to the 2D models e2ni1b1/e1ni1 and e2ni1b2/e1ni1, the continuum polarization now exhibits a plateau (roughly independent of viewing angle) during the photospheric phase, a dip near the end of the photospheric phase, followed by a sharp rise to a maximum of 1.5\% for an equator-on view at the onset of the nebular phase. Unlike for models e2ni1b1/e1ni1 and e2ni1b2/e1ni1, there is no longer a sharp peak in polarization

\section{Conclusion}
\label{sect_conc}

We have presented radiative-transfer calculations for axisymmetric Type II SN ejecta with the aim of mapping the possible landscape of continuum polarization at both photospheric and nebular epochs.
Our calculations are based on 1D radiation-hydrodynamics and 1D radiative-transfer calculations of red-supergiant star explosions and the 2D polarized radiative transfer is based on crafted 2D, axisymmetric ejecta obtained by the mapping in latitude of different 1D models. Using one reference model (compatible with the observed properties of SN\,2008bk; \citealt{lisakov_08bk_17}) for the bulk of the ejecta volume, we break spherical symmetry by placing a distinct model in a cone along the axis of symmetry. This additional model has either a larger kinetic energy, a different \nifs\ mass and distribution, or both.

Despite our limited set of simulations, our models show a broad diversity in continuum polarization at a given post-explosion epoch as well as in the evolution of that polarization with time. The continuum polarization is generally maximum at the onset of the nebular phase, as obtained previously in similar calculations \citep{DH11_pol,dessart_12aw_21}. As discussed in those works, this rise at the onset of the nebular phase results not from the unveiling of the asymmetric metal-rich inner ejecta but from the transition to a single-scattering regime in which polarization cancellation is reduced. The bulk of the asymmetry causing this large polarization in Type II SN models is located in the extended H-rich ejecta while the bulk of the radiation emanates from the \nifs-rich inner ejecta. Both absorption, scattering, and emission occur over extended regions of space, which complicates the interpretation of the polarization signatures. The maximum continuum polarization reaches several percent in models with a twice higher explosion energy  and enhanced \nifs\ mass along the poles. The observational absence of such high values in standard core-collapse SNe suggest moderate large-scale asymmetry -- extreme asymmetry in the form, for example, of jet explosions seem excluded.

Large, 0.5--1\,\% polarization at early times in the photospheric phase may be produced by an asymmetric explosion (e.g., all variants based on model e2, whose kinetic energy is twice that of the reference model e1), largely irrespective of the \nifs\ distribution but instead caused by the asymmetric density distribution. High polarization at early times is typically not observed in Type II SNe, suggesting a modest explosion asymmetry on large scale. SN\,2013ej, and more recently SN\,2021yja, stand as exceptions \citep{leonard_iauga_15,mauerhan_13ej_17,nagao_13ej_21,vasylyev_pol21yja_23}.

In contrast to observations, our simulations tend to exhibit a polarization dip at the end of the photospheric phase, just before the sudden rise to maximum. During the dip, the polarization may also flip sign, corresponding to a 90\,deg change in polarization angle. This behavior likely arises from the competition between the asymmetry of the distribution of free electrons and of the flux. Enhanced scattering occurs in regions of higher optical depth, while the bulk of the flux tends to escape through regions of lower optical depth. In general, the two distributions are opposite (i.e., if the former is prolate, the latter is oblate, and vice versa). This feature may be exacerbated by our idealized 2D  setup. In nature, the inner ejecta of Type II SNe may be asymmetric mostly on small scale, in the sense that any large scale asymmetry may be destroyed by the numerous fluid instabilities taking place during and after the explosion. For example, the \nifs\ bubble effect acting over weeks may eventually completely destroy any clean inner-ejecta asymmetry present immediately after the explosion (say 1 minute after core bounce).

Optical depth effects are also found to persist until very late times, even in the nebular phase. This not only affects the SN brightness as seen by a distant observer along different viewing angles, but it also leads to a peculiar evolution of the continuum polarization. For example, it often deviates from a $1/t^2$ fall-off, or may even rise and peak with a delay during the nebular phase. Although clearly limited by the parameter space explored, the ``plateaus'' in nebular-phase continuum polarization are obtained exclusively by models with enhanced inner-\nifs\ only (i.e., the models e1[2]ni2/e1ni1). The presence of isolated \nifs\ blobs may also induce peculiar evolution patterns for the continuum polarization, since they may maintain locally a high ionization and optically-thick conditions when the rest of the ejecta is optically thin. Such ionization shifts are likely important in Type II SNe because the roughly 100\,d-long plateau phase is fundamentally tied to recombination: without recombination, the photospheric phase of Type II SNe would last for about two years.

The late time evolution may also be affected by our assumption of a smooth ejecta. Clumping would hasten the recombination of the ejecta (even during the photospheric phase; \citealt{d18_fcl}) and might hasten the drop of the continuum polarization during the nebular phase. At such times, a better treatment of chemical mixing is also warranted \citep{jerkstrand_04et_12,DH20_shuffle} so our models are not as robust as one would desire but the standard boxcar mixing was applied here, as in the similar \cmfgen\ models produced for SN\,2008bk by \citet{lisakov_08bk_17}, in order to keep the computational costs reasonable. Future work should be based on ejecta obtained with multi-D explosion simulations (either 2D ejecta or 3D ejecta exhibiting a dominant axis of symmetry) in order to improve the physical consistency and realism of our polarization calculations.

\begin{acknowledgements}
D.C.L. acknowledges support from NSF grants AST-1009571, AST-1210311, and AST-2010001, under which part of this research was carried out. D.J.H. thanks NASA for partial support through the astrophysical theory grant 80NSSC20K0524.  This work was granted access to the HPC resources of  CINES under the allocation 2020 -- A0090410554 and of TGCC under the allocation 2021 -- A0110410554 made by GENCI, France. This research has made use of NASA's Astrophysics Data System Bibliographic Services.
\end{acknowledgements}

\appendix

\section{Polarized radiative transfer with \longpol}
\label{sect_nomenclature}

For completeness, we summarize the nomenclature and sign conventions adopted in \longpol\ and also presented in \citet{DH11_pol}.  We assume that the polarization is produced by electron scattering. The scattering of electromagnetic radiation by electrons is described by the dipole or Rayleigh scattering phase matrix. To describe the ``observed'' model polarization we adopt the Stokes parameters  $I$, $Q$, $U$, and $V$ \citep{Cha60_rad_trans}. Since we are dealing with electron scattering, the polarization is linear and the $V$ Stokes parameter is identically zero. For clarity, $I_Q$ and $I_U$ refer to the polarization of the specific intensity, and $F_Q$ and $F_U$ refer to the polarization of the observed flux.

For consistency with the  earlier work of \cite{hillier_94, hillier_96} we choose a right-handed set of unit vectors $(\zeta_X,  \zeta_Y, \zeta_W)$. Without loss of generality the axisymmetric source is centered at the origin of the coordinate system with its symmetry axis lying along $\zeta_W$, $\zeta_Y$ is in the plane of the sky, and the observer is located in the XW plane.

We take $F_Q$ to be positive when the polarization is parallel to the symmetry axis (or more correctly parallel to the projection of the symmetry axis on the sky), and negative when it is perpendicular to it. With our choice of coordinate system, and since the SN ejecta are left-right symmetric about the XW plane,  $F_U$ is zero by construction. This must be the case since symmetry requires that the polarization can only be parallel, or perpendicular to, the axis of symmetry. For a spherical source, $F_Q$ is also identically zero.

$I(\ip,\delta)$, $I_Q(\ip,\delta)$ and $I_U(\ip,\delta)$ refer to the observed intensities on the plane of the sky. $I_Q$ is positive when the polarization is parallel to the radius vector, and negative when it is perpendicular. In the plane  of the sky we define a set of polar coordinates ($\ip,\delta$)  with the angle $\delta$  measured anti-clockwise from $\zeta_Y$. The polar coordinate, $\ip$, can also be thought of as the impact parameter of an observer's ray. We also use the axes defined by the polar coordinate system to describe the polarization. $F_I$ is obtained from $I(\ip,\delta)$ using

\begin{equation}
F_I= {2 \over d^2}  \int_0^{\ip_{\rm max}} \int_{-\pi/2}^{\pi/2}  \, I(\ip,\delta) dA \, ,
\end{equation}

\noindent
where $dA=\ip d\delta d\ip$. Since $\zeta_\ip$ is rotated by an angle $\delta$ anticlockwise from $\zeta_Y$,
$F_Q$ is given by

\begin{equation}
F_Q = {-2 \over d^2} \int_0^{\ip_{\rm max}}  \int_{-\pi/2}^{\pi/2}
\left[ I_Q(\ip,\delta)  \cos 2\delta +  I_U(\ip,\delta)   \sin 2\delta \right] \, dA \,.
\end{equation}

\noindent
In a spherical system, $I_Q$ is independent of $\delta$, and $I_U$ is identically zero.

\noindent
In this study, we focus on the polarization in the relatively line-free region bounded by 6900 and 7200\,\AA\ and refer to this as ``continuum'' polarization. In practice, we quote average values over that spectral region. We either discuss the percentage continuum polarization $<P_{\rm cont}>$ defined as $100  | F_Q \, \big / F_I|$  or $<Q_{\rm cont}>$ defined as $100 F_Q \, \big / F_I$.

\section{Additional illustrations}
\label{sect_appendix}

\begin{figure*}
\epsfig{file=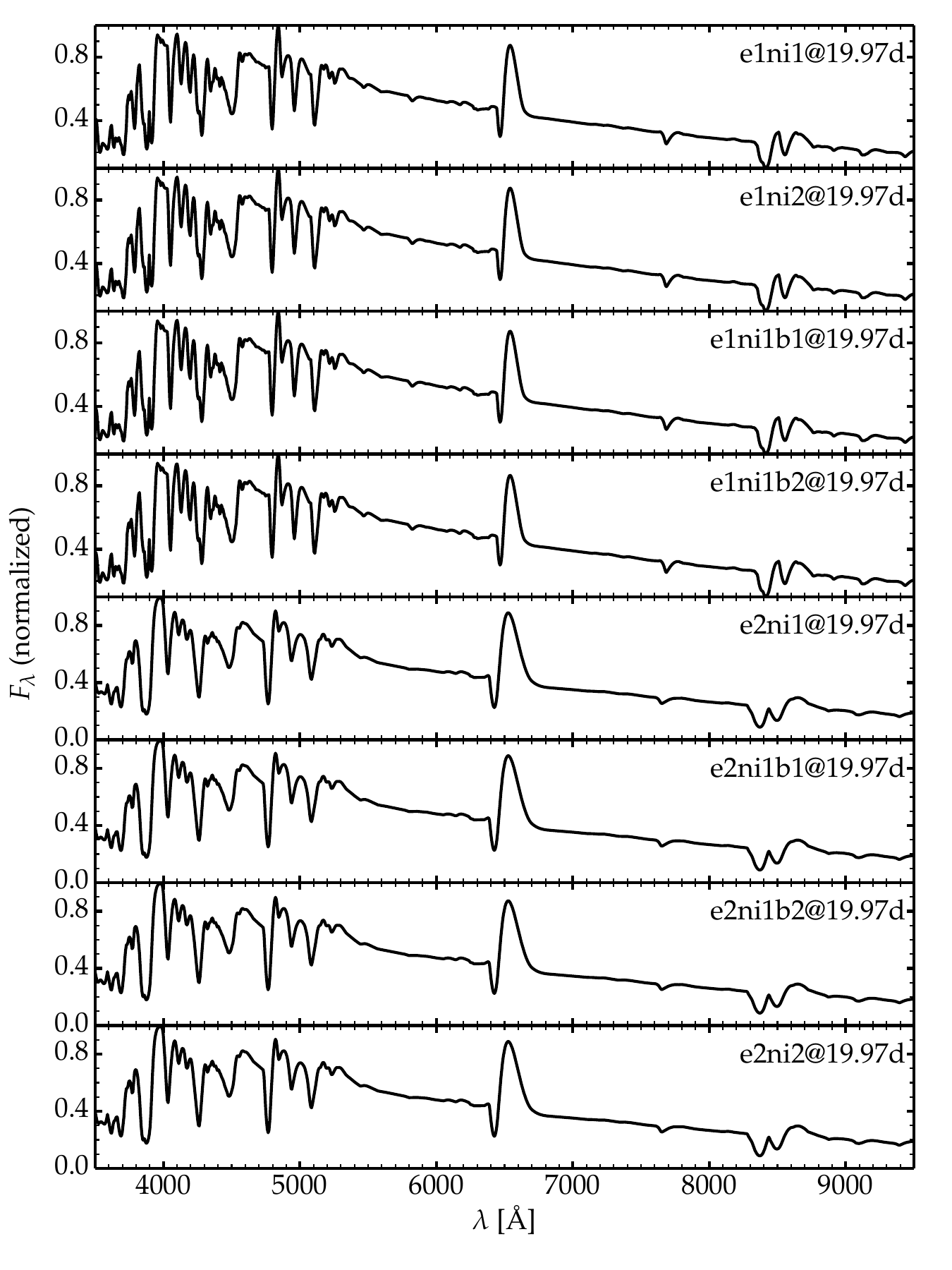, width=9.2cm}
\epsfig{file=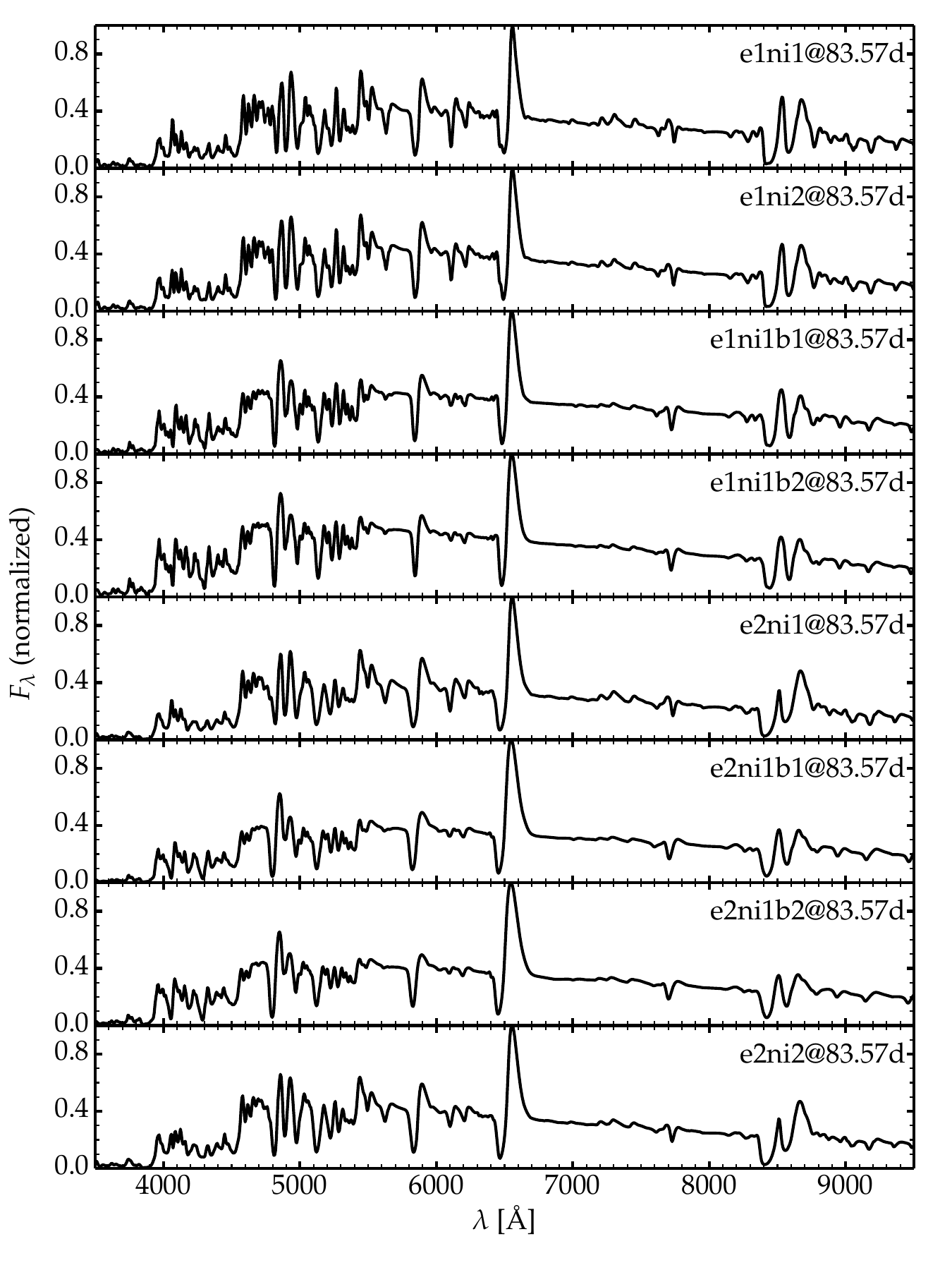, width=9.2cm}
\caption{Spectral comparison between 1D \cmfgen\ models at about 20 (left) and 84\,d (right) after explosion.
  \label{fig_spec_1d_comp}
}
\end{figure*}

\begin{figure*}
\begin{center}
\epsfig{file=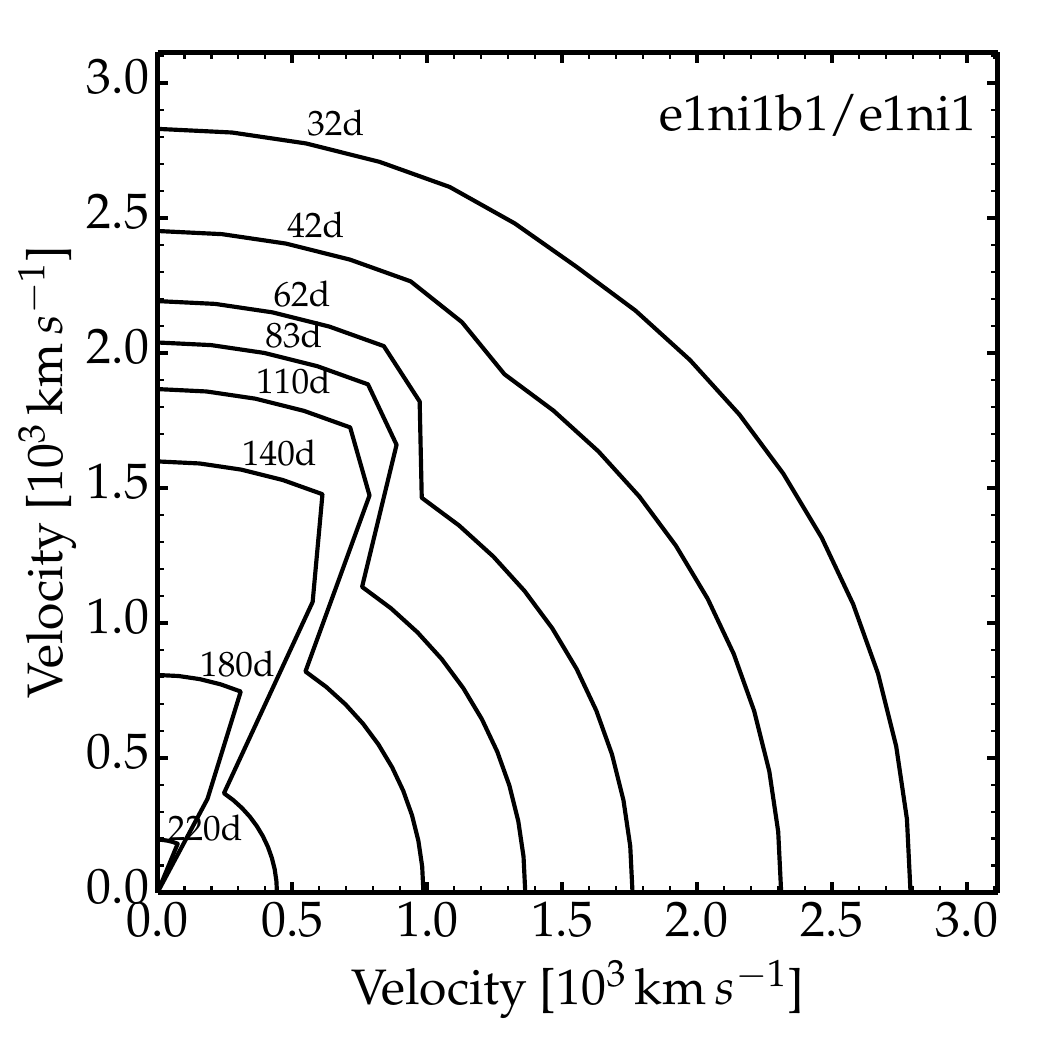, width=7.5cm}
\epsfig{file=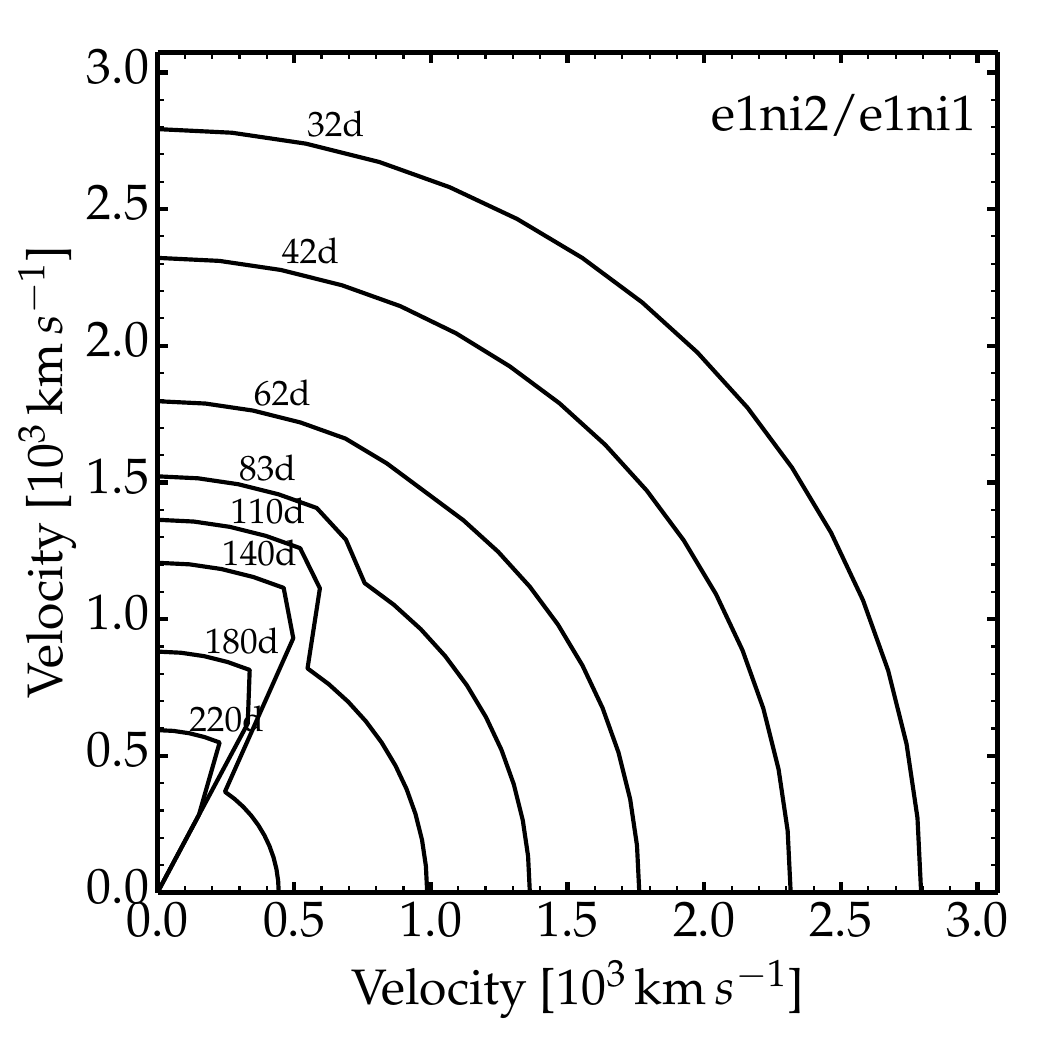, width=7.5cm}
\epsfig{file=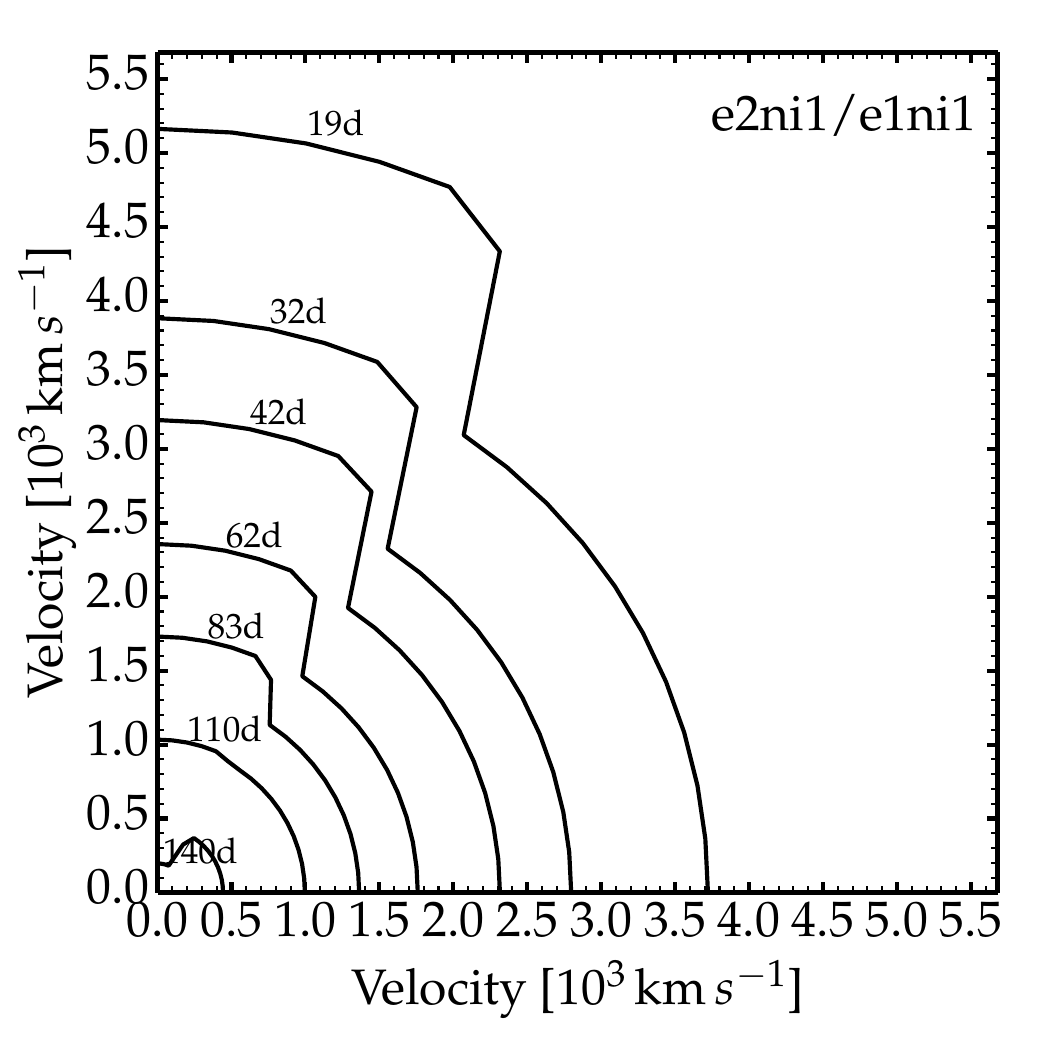, width=7.5cm}
\epsfig{file=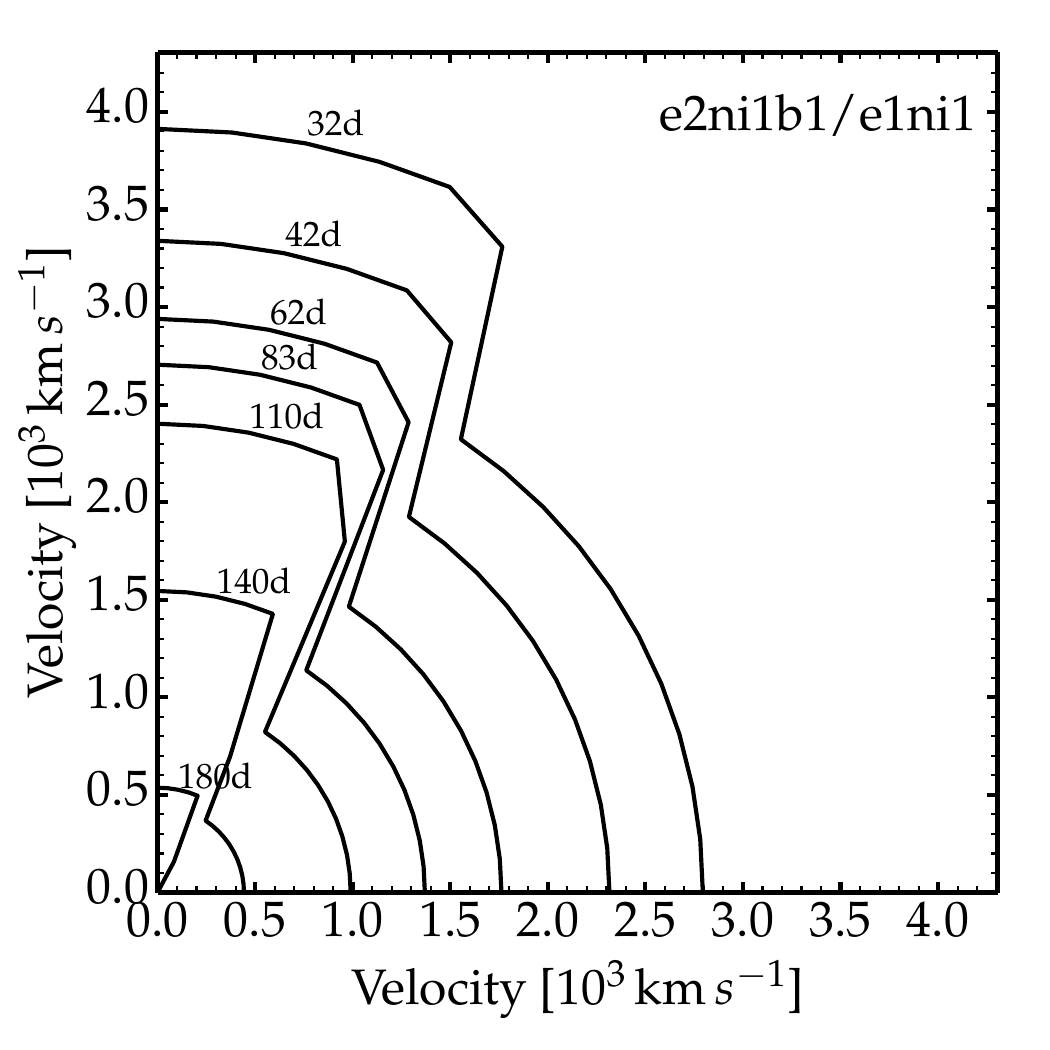, width=7.5cm}
\epsfig{file=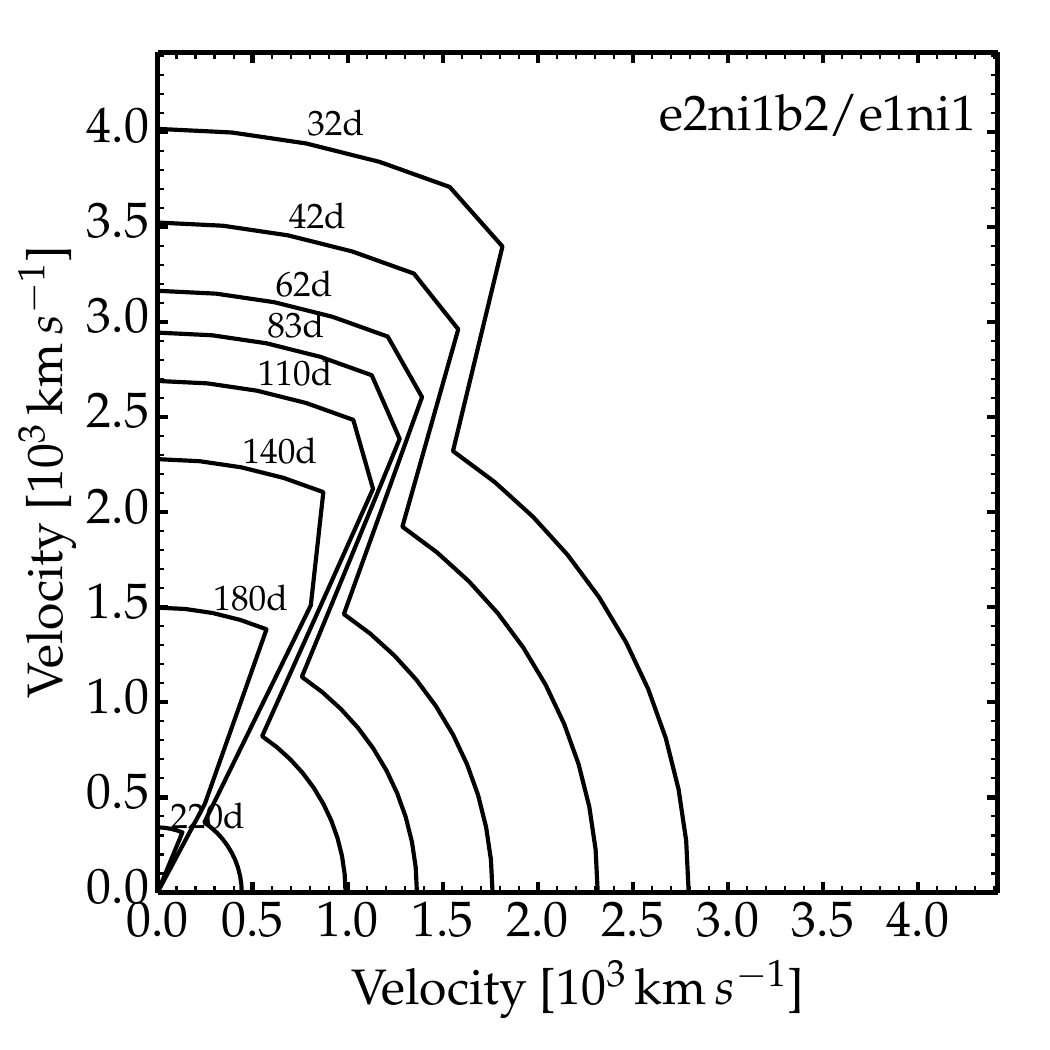, width=7.5cm}
\epsfig{file=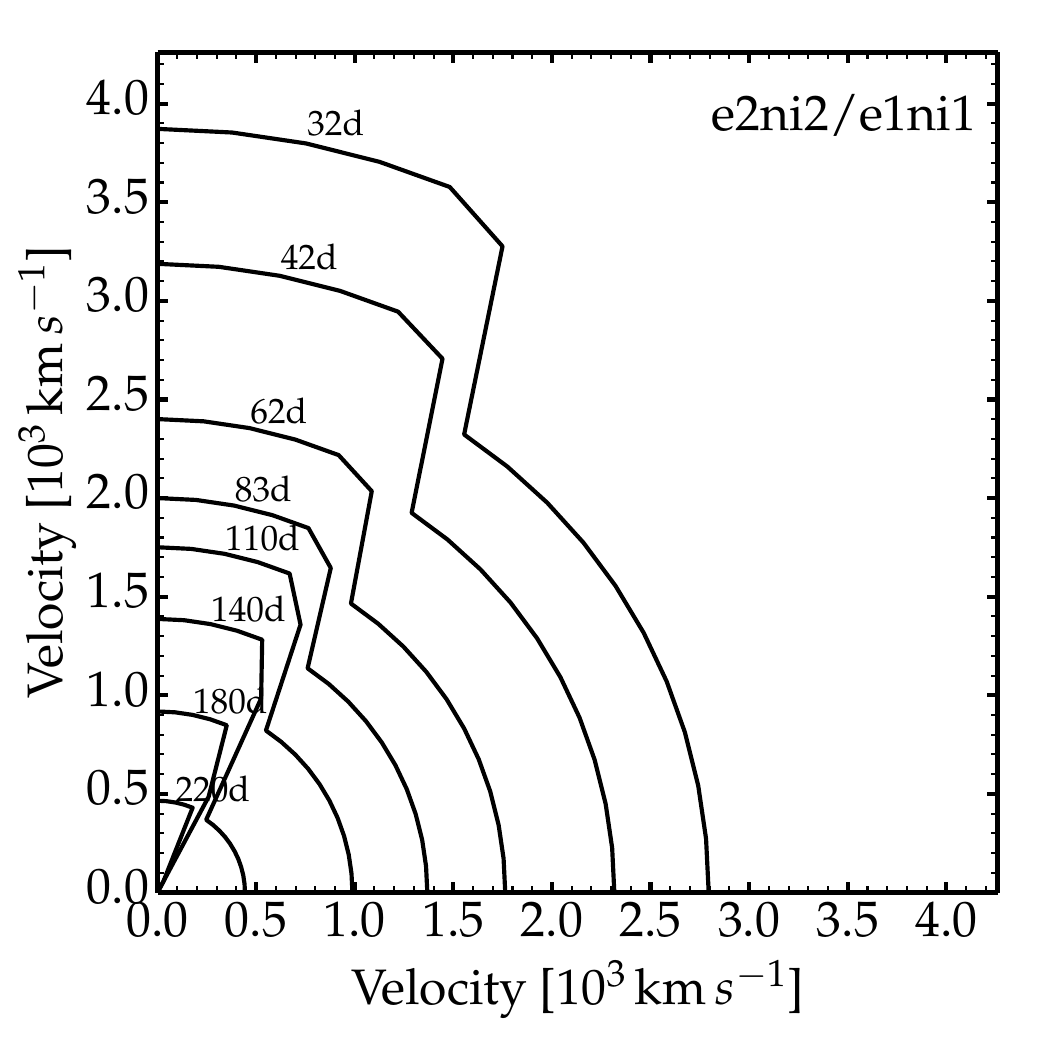, width=7.5cm}
\caption{Same as Fig.~\ref{fig_2d_phot} for model e1ni1b2/e1ni1 but showing the evolving morphology of the electron-scattering photosphere for the rest of the 2D model set.
\label{fig_2d_phot_rest}
}
\end{center}
\end{figure*}

\begin{figure*}
\begin{center}
\epsfig{file=e1ni1b2_e1ni1_lc_mv_pcont.pdf, width=9cm}
\epsfig{file=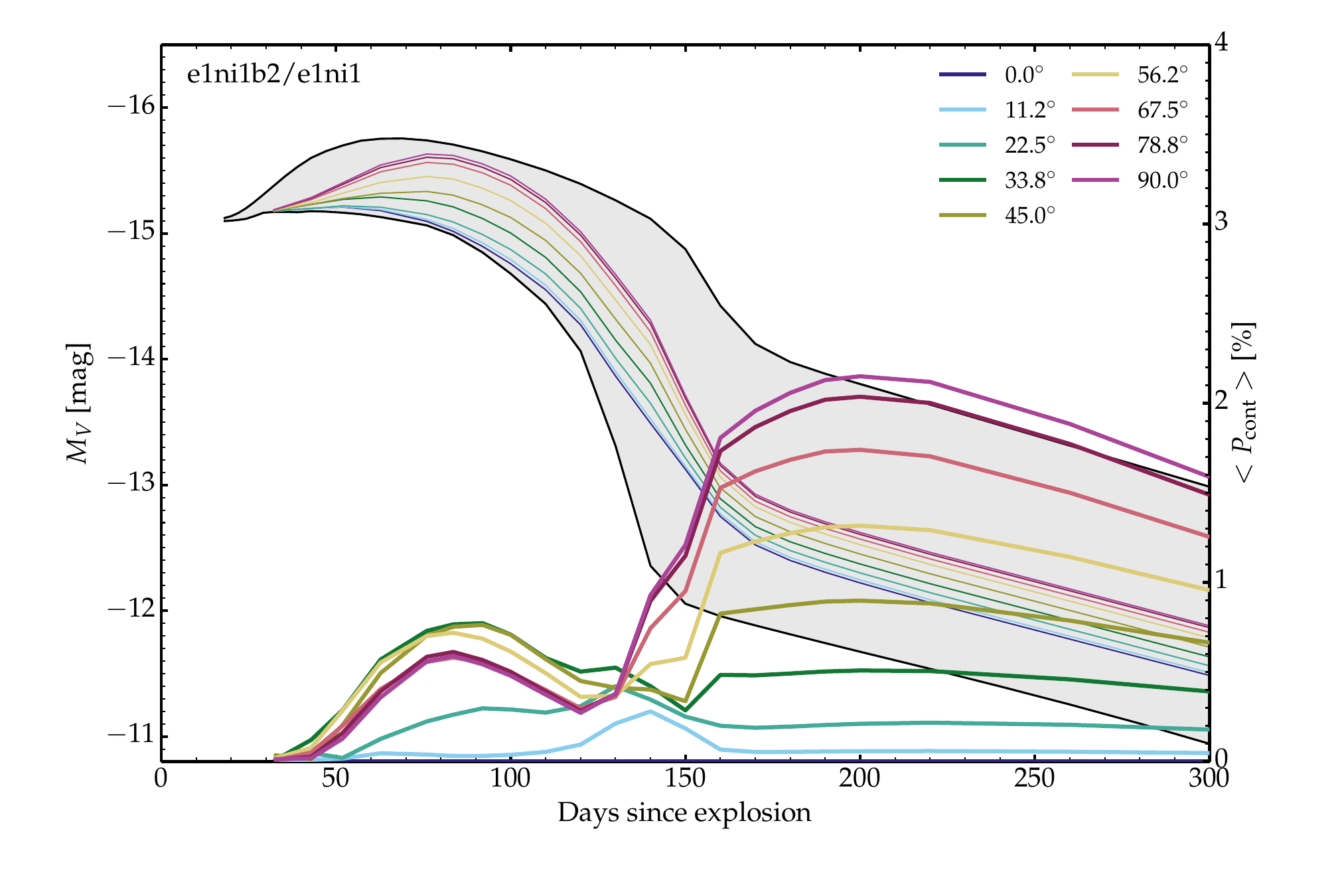, width=9cm}
\caption{Influence of the spectral range used for computing the average continuum polarization. In the left panel, the range covers from 6900 to 7200\,\AA\ and in the right panel, the range covers from 6900 to 8200\,\AA.
\label{fig_cont_range}
}
\end{center}
\end{figure*}

\begin{figure*}
\begin{center}
\epsfig{file=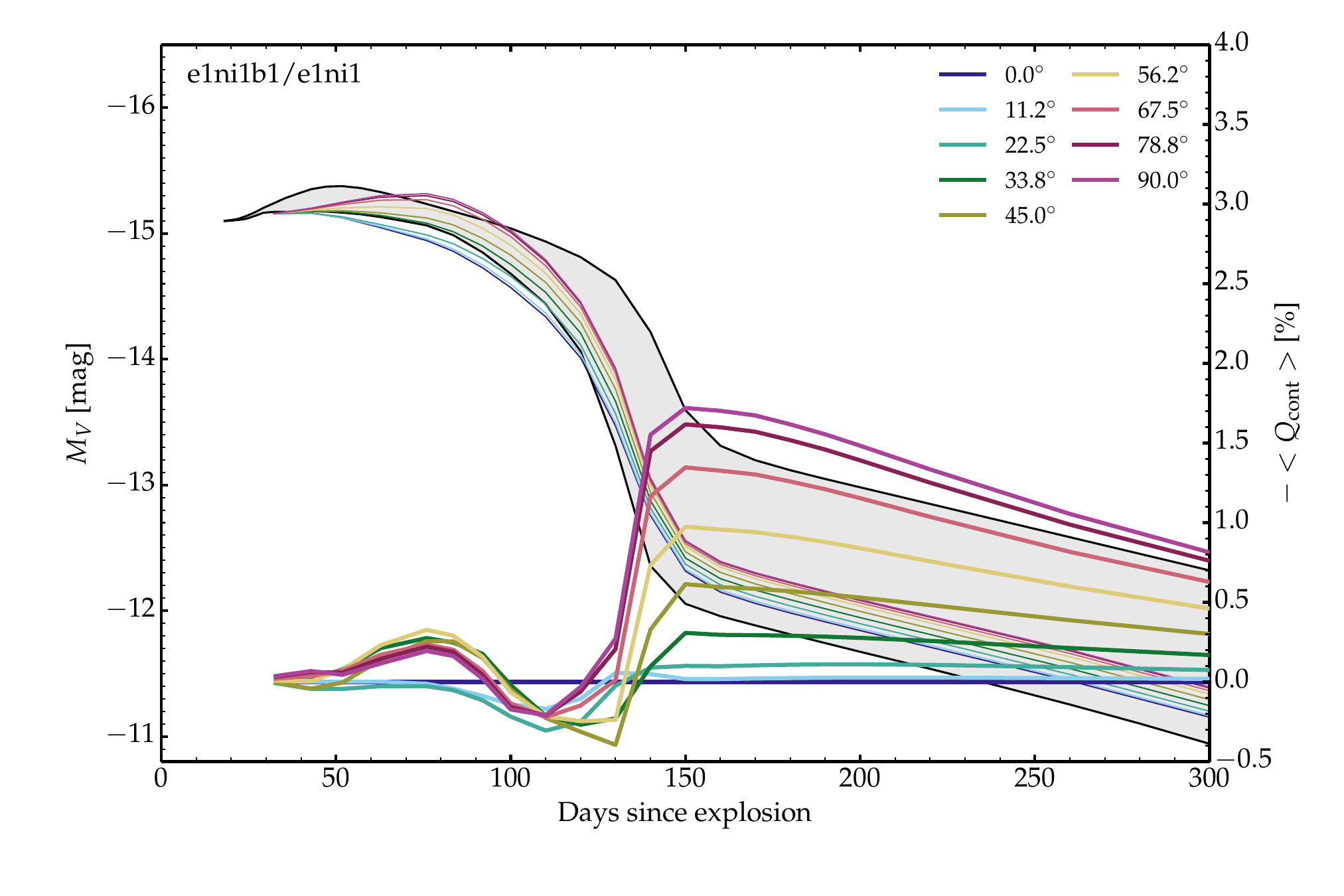, width=8cm}
\epsfig{file=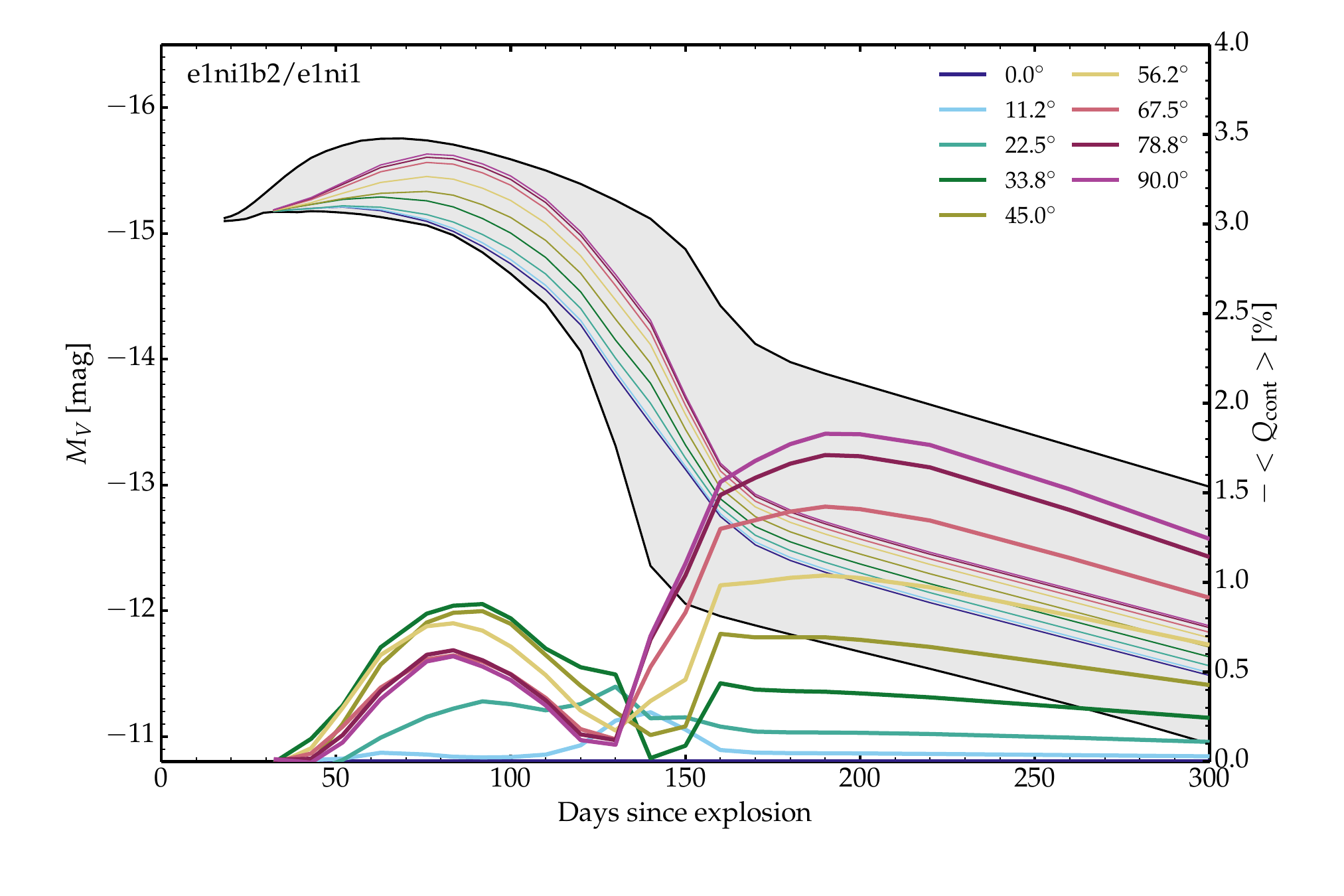, width=8cm}
\epsfig{file=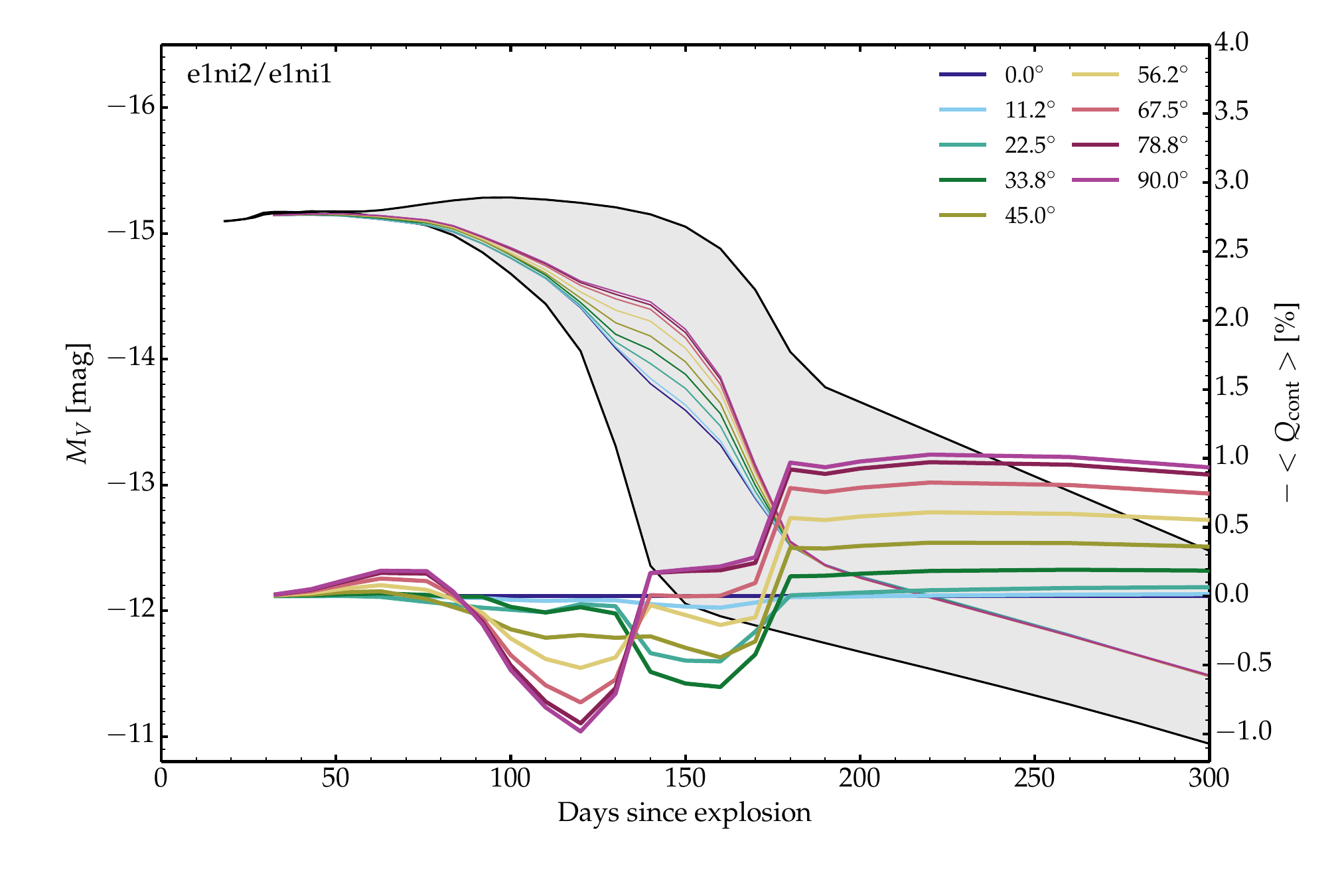, width=8cm}
\epsfig{file=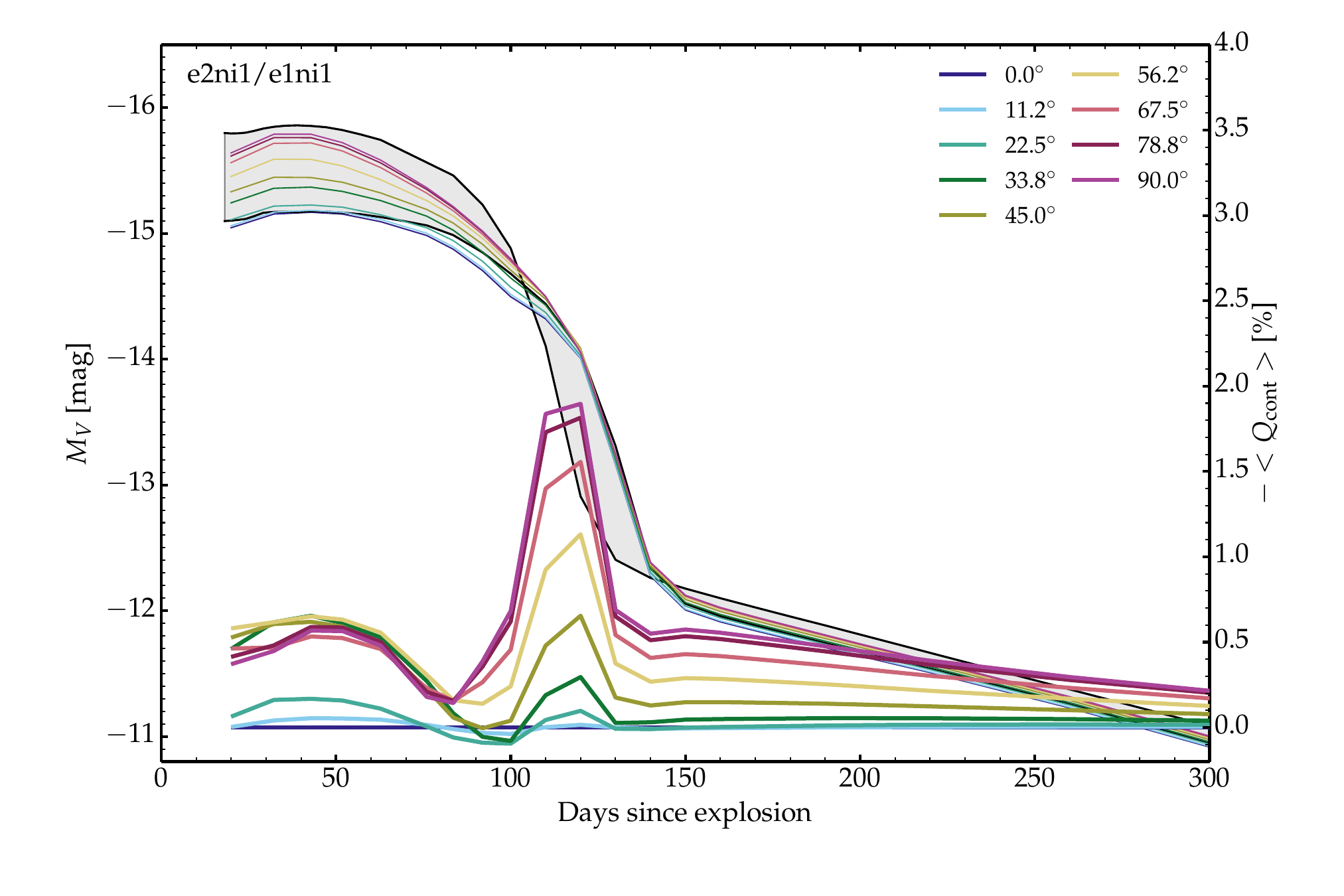, width=8cm}
\epsfig{file=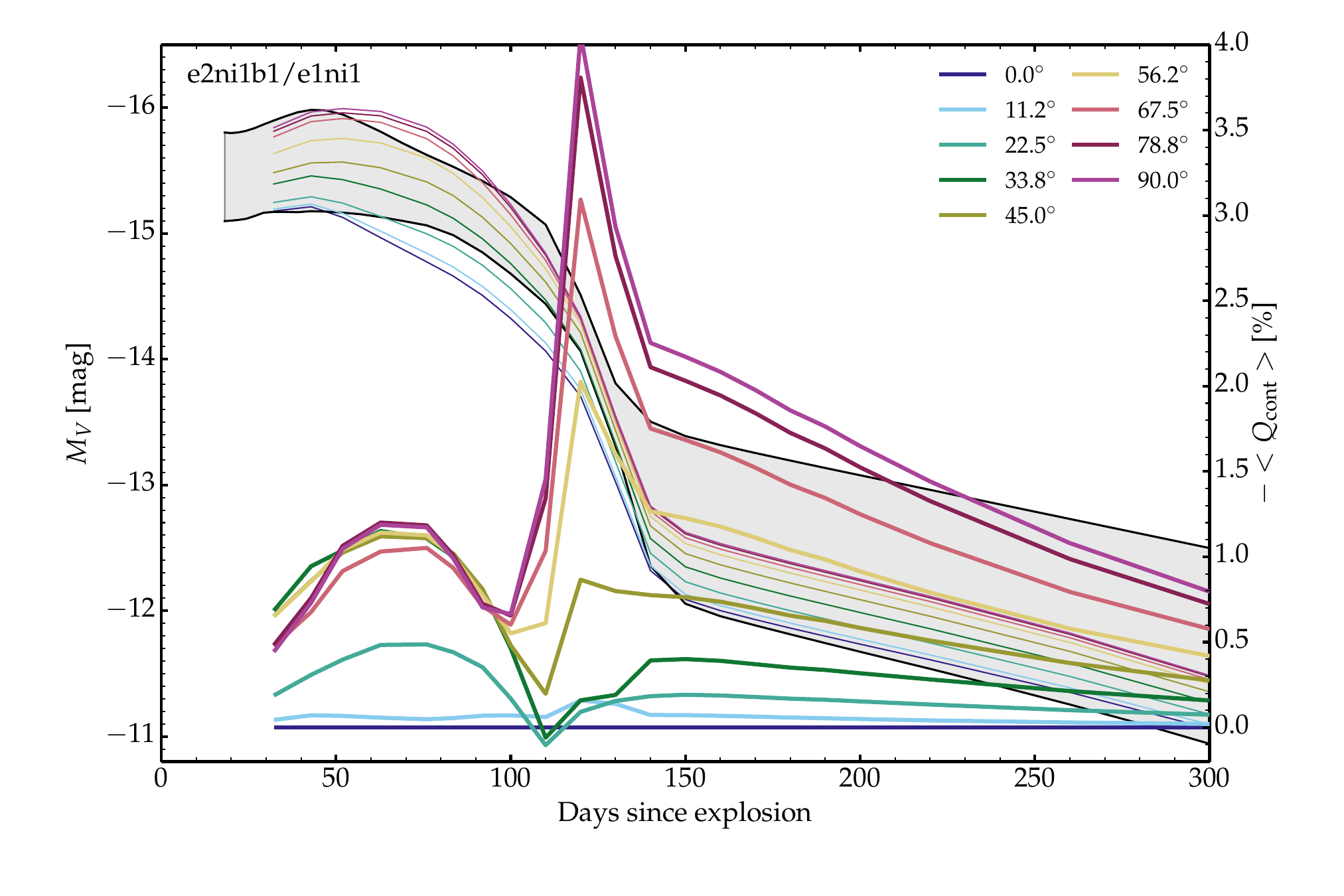, width=8cm}
\epsfig{file=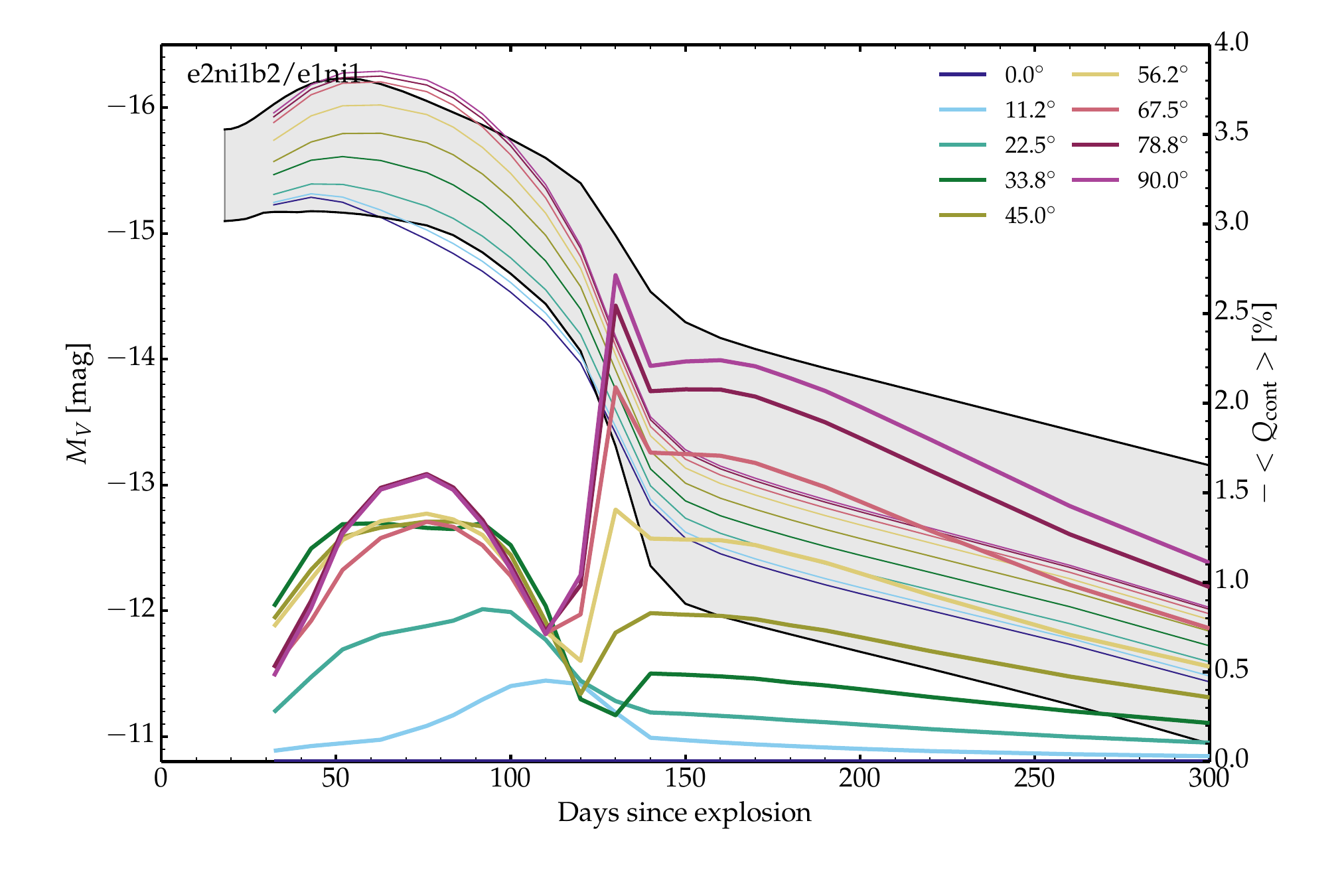, width=8cm}
\epsfig{file=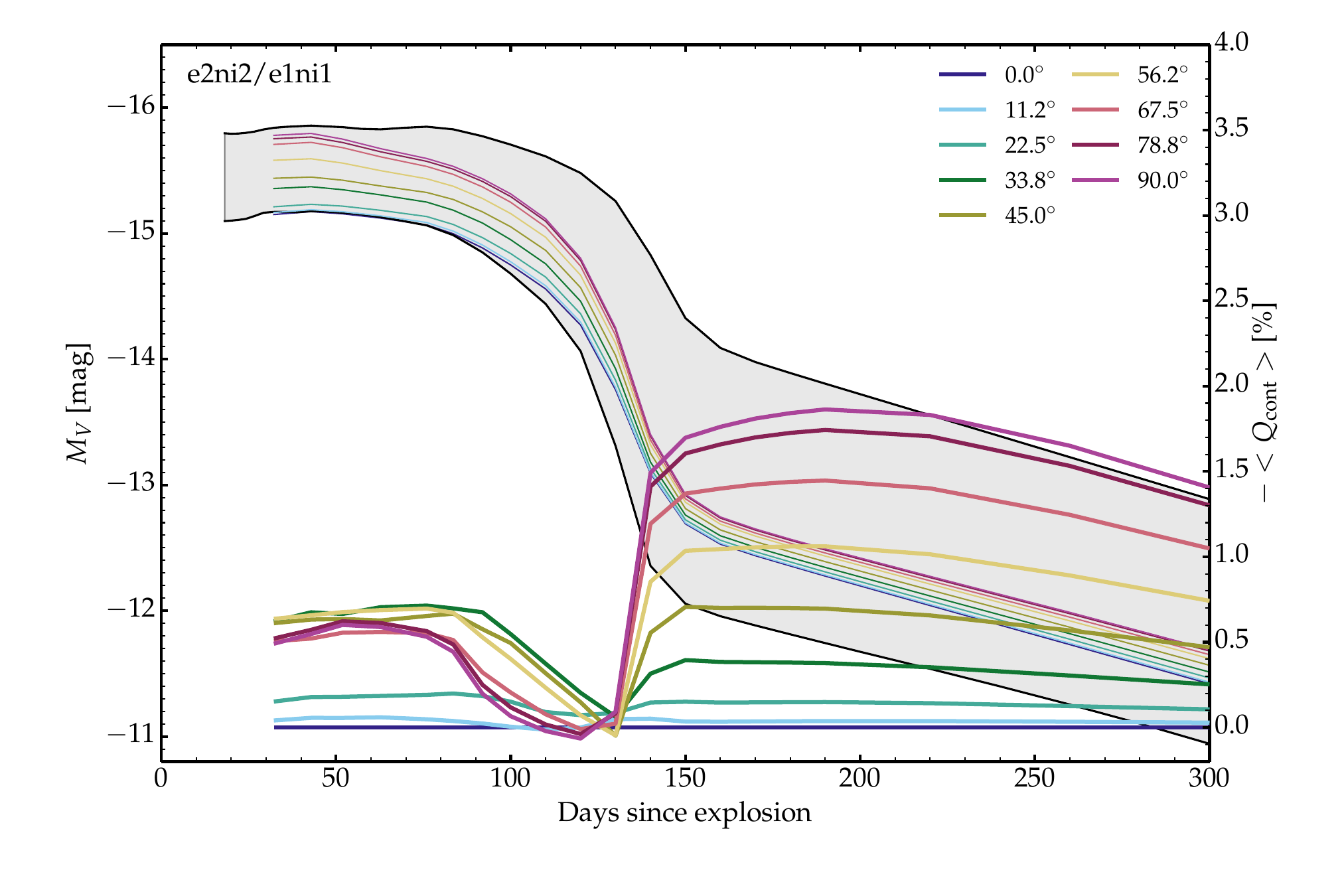, width=8cm}
    \caption{Same as Fig.~\ref{fig_e1ni1b2_e1ni1} but showing the quantity $<-Q_{\rm cont}>$ (averaged over the spectral region extending from 6900 until 7200\,\AA), which reveals any potential polarization sign flip (unlike $P_{\rm cont}$). In the nomenclature of \citet{dessart_12aw_21}, this quantity is defined as $-100\,F_Q/F_I$ (see Appendix~\ref{sect_nomenclature}). We show the negative of $Q_{\rm cont}$ so that most of the polarization values are positive. With our sign convention, a negative $F_Q$ corresponds to an electric vector perpendicular to the symmetry axis.
\label{fig_qcont}
}
\end{center}
\end{figure*}

\end{document}